%% file: main.tex
\pdfoutput=1

\documentclass[10pt]{article}
\usepackage{amsmath}
\usepackage{amsfonts}
\usepackage{amssymb}
\usepackage{multicol}
\usepackage{graphicx}
\usepackage{subfigure}
\usepackage{url}

\input{tcilatex}
\begin{document}

\input{klim_tit}



\section{Introduction}

This work is dedicated to promoting and expanding Boltzmann's hypothesis
linking the perceived direction of time to the second law of thermodynamics 
\cite{Boltzmann-book}. Ideas of Ludwig Boltzmann created many disputes and
controversies during his life, but, nevertheless, left us a rich inheritance
that forms the cornerstone of modern science. Objectively, many statements
of Boltzmann lacked rigor and this, in combination with the traditional
disbelief of old science in reality of molecules, invited now century-old
criticism of his work. At the same time, Boltzmann's ideas displayed deep
intuition that was fully understood and appreciated by his followers (who
include such figures as Planck and Einstein) only after his death.
Boltzmann's time hypothesis is one of these great ideas,\ whose importance
could not possibly be understood, let along appreciated, during his times.
Yet, the world is slowly moving towards the understanding of the perceived
direction of time offered by Boltzmann more than a century ago.

There is a great number of publications discussing the direction of time
from the perspectives of philosophy and physics. The topic is virtually
infinite and the reader can be referred to the principal publications: books 
\cite{Russell2009,Reichenbach1971,PriceBook,Time1997} on the philosophical
side and books \cite{Prig1980,PenroseBook,Zeh2007,Hawking-time2011}\ on the
physical side of the argument about the direction of time. In this respect,
the present work offers only a brief introductory narrative explaining some
preliminaries. The focus of this work is Boltzmann's view on the perceived
direction of time, its implications and, most importantly, the possibility
of experimental testing of the time mechanism, which is opened by the
Boltzmann hypothesis. We pay attention to the Diosi-Penrose theory \cite%
{PenroseBook} suggesting that thermodynamic time can be induced by gravity,
but note that this theory is just one of many possibilities. In the context
of the Boltzmann time hypothesis, the microscopic effect of "flowing" time
can be detected without assuming any specific time-generating mechanism,
although this requires rare cases in particle physics known as CP violations 
\cite{PDG2012}. A number of terminological and methodological issues are
discussed in the Appendix.

\ 

\section{Intuitive perception of time}

Perception of time is an inseparable part of our consciousness. Humans
recognise their existence as individuals in the context of their awareness
of time. This recognition, which we may call self-awareness, is focused on
the present moment of time -- the moment through which the past moments
known to us are continuously appended by events from as-yet unknown future.
Our intuitive perception of time is perhaps best described by the lines

\begin{quote}
That river of time keeps on flowing

The present turns into the past
\end{quote}

\noindent taken from poem \textquotedblleft The river of
time\textquotedblright\ by Linda Ori. These poetic words give an accurate
metaphoric account of our intuitive perception of time. This perception can
be expanded into the following six principles of Hans Reichenbach \cite%
{Reichenbach1971}:

\begin{enumerate}
\item Time goes from past to the future

\item The present which divides the past from the future is now

\item The past never comes back

\item We cannot change the past but we can change the future

\item We can have records of the past, but not of the future

\item The past is determined, the future is undetermined
\end{enumerate}

\noindent These principles would be considered by most people as being
obvious, which reflects the strength of our intuitive perception of time.
Yet, it is arguable that the perceived flow of time is only a metaphoric
description of the reality encompassing groups of dependent events that
follow each other in time. Zwart \cite{Zwart1972} analysed this problem and
remarked:

\begin{quote}
There is no flow of time beside or beneath the flow of events, but the flow
of time is nothing but the flow of events.
\end{quote}

\noindent Hence, the flow of time is represented by chains of ordered
events, while each of these events is associated with a certain time moment
and creates conditions for the following event(s) to appear. The directions
along this chain --- the past and the future --- are obviously not
equivalent: we think that the future events depend on past events but not
vice versa. This non-equivalence of two temporal directions (toward the past
and toward the future) is called the arrow of time. In the context of this
non-equivalence, we often say that the preceding events cause the following
events. This leads us to concept of causality, which we understand very well
at the intuitive level and expect that, with the use of causality, a more
rigorous and justified account can be provided for our intuitive time
experience. This view was articulated by Reichenbach \cite{Reichenbach1971}
more than half-a-century ago:

\begin{quote}
Casual connection is a relation between physical events and can be
formulated in objective terms. If we define time order in terms of casual
connection, we have shown which specific features of physical reality are
reflected in the structure of time, and we given an explication to the vague
concept of time order.
\end{quote}

This program of expressing our basic intuitive perception of time in terms
of most fundamental casual relationship between events from the past,
present and future is, generally, a very constructive idea, which has been
followed by many scholars. Without invoking physical laws, however, the
philosophical program outlined by Reichenbach encounters mounting
difficulties (briefly discussed below) and was not able to produce a
transparent and concise representation of our temporal intuition in terms of
objective casual connections.

\section{Universal causality}

Existence of causality in its broadest sense is often seen as being a most
fundamental property of the universe, a priori postulate that does not need
any further justification of explanation. As Bertrand Russell wrote in his
book "Our knowledge of the external world" \cite{Russell2009}

\begin{quote}
The law of causation, according to which later events can theoretically be
predicted by means of earlier events, has often been held to be a priori, a
necessity of thought, a category without which science would not be possible.
\end{quote}

\noindent We stress that this understanding of temporal causation, which we
can call "universal causality", is a philosophical (rather than scientific)
concept. We are interested only in causality associated with the direction
of time (e.g. the past events causing the future events) and avoid any other
interpretations of causality, which might also be important in philosophy.
The domains of philosophy and science may overlap but, generally, do not
coincide. The former reflects our most fundamental perceptions of reality,
while the latter involves hypotheses and theories that can be tested
experimentally and confirmed or invalidated. The principle of universal
causality is beyond testable physical laws as these laws are deemed to be
based on this principle. This, however, does not imply that universal
causality is unconditionally supported in philosophy and objected in physics
(in fact the opposite might be more correct: many philosophers refer to
physical laws to clarify the concept of causality, while physicists tend to
use causality as intuitive a priori to explain physical laws). The intuition
based on causality is deeply embedded into human consciousness and most
scientists tend to follow this intuition. Philosopher Huw Price \cite%
{PriceBook} writes:

\begin{quote}
I have been trying to correct a variety of common mistakes and
misconceptions about time in contemporary physics --- mistakes and
misconceptions whose origins lie in the distorting influence of our own
ordinary temporal perspective, and especially of the time asymmetry of that
perspective.
\end{quote}

As noted above, experimental validation of universal causality is
conceptually problematic. Even defining the meaning of "something causing
something else" without invoking various physical laws is not a simple
matter. For example, consider two events shown in figure \ref{fig1}

\begin{description}
\item A.\qquad the vase falls

\item B.\qquad the vase breaks into pieces
\end{description}

\noindent Assume that the floor is hard (so that A inevitably causes B) and
there is no malicious hooligan who can break the vase by some other means
(so that B is impossible without A). While our intuition unmistakably
identifies A as the cause and B as an effect, this intuition is not easy to
define at a general level without appealing to physical laws that, in fact,
may be implicitly based on causality. Obviously A and B are linked to each
other. There is, however, no logical flaw in identifying B as the cause and
A as the effect despite the fact that this contradicts our intuition: given
that the vase is shattered into pieces, it must have fallen from the table.
This basic example illustrates difficulties in formulating a broad
definition of causality that involves reversible, irreversible and
stochastic processes of microscopic and macroscopic scales and agrees with
our intuitive perception of time. Russell \cite{Russell2009} noted that

\begin{quote}
The view that the law of causality itself is a priori cannot, I think, be
maintained by anyone who realises what a complicated principle it is.
\end{quote}

Even more recent publications \cite{Dowe1992,Time1997} have not brought a
complete resolution to this problem. One can, of course, invoke the second
law of thermodynamics, which, for the case illustrated in figure \ref{fig1},
determines that state A of the vase must precede state B. The cause and the
effect can then be distinguished among casually connected events by their
temporal precedence, defining the earlier event as the cause and the later
event as the effect \cite{Dowe1992}. Generally, this definition is
reasonable but, in the present context, this leads us to a circular argument
--- the direction of time is determined by causality, while causality is
defined by temporal precedence. Defining causality in terms of precedence is
perfectly acceptable from the philosophical perspective, but this definition
downgrades universal causality to being a consequence of temporal
precedence, which, in its turn is constrained by the laws of physics (i.e.
by the second law of thermodynamics). In this context, we can simply
overlook temporal precedence and define, for related events A and B, a lower
entropy state (state A) as the cause and a higher entropy state (state B) as
the effect. It seems that linking causality to irreversibility has become
commonly accepted in modern philosophy and is sometimes referred to as the
"standard interpretation"\cite{Time1997}. This leads us to a less universal
interpretation of causality that is connected to the physical laws, instead
of being an unquestionable foundation of these laws.

Some recent experiments \cite{Delayed-choice-2015}, which are based on
Wheeler's idea of delayed choice in quantum measurements, are sometimes seen
as action of retro-causality in the quantum world. Considering the
difficulty of finding a rigorous and comprehensive definition of causality,
these experiments can only indicate that the concept of causality is less
useful in interpretation of the microworld than in interpretation of the
macroworld and, perhaps, less universal than it was previously thought.
While results of these experiments may be alarming for proponents of
universal causality, the usefulness of a more practical understanding of
causality is not likely to be affected.

\section{"Practical" causality\label{SecP}}

Consider an engineer who needs to simulate an industrial process over a
given time interval. The engineer is hesitant whether he/she should set
initial or final conditions for the process. Based on his/her intuition or
explicitly using the causality principle (the future is determined by the
past) the engineer decides to set the initial conditions and is, generally,
absolutely right to do this. The engineer's concern is to obtain the most
accurate simulations with minimal effort while bypassing complicated
problems related to the nature of time. Invoking temporal causality and
setting the initial conditions is, typically, the shortest path to practical
success. The causality principle works quite well in most practical cases,
replacing complex analysis of the properties of time by a simple rule of
thumb: giving preference to the initial conditions over the final conditions.

There might be some cases when any type of conditions (initial or final)
would allow us to perform the simulations equally well. The problem of
elastic collisions of two particles may serve as an example. There are also
cases when neither condition can guarantee a success. For example, evolution
of a complex stochastic system is not well-defined by either initial or
final conditions. However, the engineer's decision is absolutely reasonable,
as practical cases in which setting the final conditions is better than
setting the initial conditions are hard to find. Let us consider few
examples of such rare cases. 1) It is easier to use the final conditions
than initial conditions in the process of deciphering a message, 2)
determining the path of a tornado might be easier using final conditions
when the swath of plants and constructions destroyed by the tornado is known
and 3) evaluating an experiment is easier after its completion, when the
experimental records are available. We note that in all of these cases the
apparent entropy associated with the information of interest is reduced
forward in time (which is, of course, compensated by a large increase of
entropy somewhere else). The last example involves continuous human
interference recording most valuable experimental information; its complete
simulation may need a combination of the initial conditions (i.e. the
experimental set up) and final conditions (i.e. the recorded results).

Practical causality has a broad range of applications. Engineering is
mentioned to emphasise the applied character of the interpretation of
causality considered in this section. Besides engineering, this
interpretation is also used commonly and effectively in various fields of
science. The intuitive use of causality is justified as long as we do not
need to dwell on investigating the fundamental properties of time and
explain causality instead of presuming it --- practical causality is a very
useful concept and, generally, it is consistent with our experiences.

\section{Boltzmann's time hypothesis\label{SecB}}

The arguments of the previous sections, outlining the utility of practical
causality and difficulties of interpreting causality as a priori standing
above and beyond physical laws, lead us to declare an epistemological
connection between causality and the laws of physics.\ This section makes
the next logical step, adding a direct physical relation to the previously
declared connection: causality is to be understood as a consequence the
physical laws and, hopefully, can be tested in conjunction with these laws.

While the directional nature of time needs to be related to measurable
physical quantities, establishing this relation seems problematic --- all
major physical theories, including classical and quantum mechanics, special
and general relativity are time-symmetric. There is an important exception
represented by thermodynamics: entropy tends to increase forward in time
according to the second law of thermodynamics. As previously noted, the
directional properties of the second law have been repeatedly used by
philosophers to define the preferential direction of time \cite{Dowe1992}.
This treatment, however, does not necessarily represent a physical
hypothesis, which requires not only correspondence but also physical
equivalence of the perceived flow of time and the direction of entropy
increase.

The physical time hypothesis was introduced by Boltzmann, who explains in
his "Lectures on gas theory" \cite{Boltzmann-book}:

\begin{quote}
However, just as at a particular place on the earth's surface we call "down"
the direction toward the center of the earth, so will a living being in a
particular time interval of such a single world distinguish the direction of
time toward the less probable [i.e. lower entropy --AYK] state from the
opposite direction (the former toward the past, the latter toward the
future).
\end{quote}

\noindent In his remarkable insight, Boltzmann sees the perceived direction
of time as being physically determined by the temporal direction of the
entropy increase: if entropy had to decrease in some segments of the
universe, people living there would see our future as their past and our
past as their future. Reichenbach \cite{Reichenbach1971} and Hawking \cite%
{Hawking-time2011} believe that the thermodynamic interpretation of the
arrow of time explains existence of the psychological time arrow: we
remember the past but do not remember the future. This point can be
illustrated by the example "footsteps on a beach" shown in figure \ref{fig2}
--- do these footsteps indicate that someone has walked on the beach in the
past or that someone will walk on the beach in the future? While the answer
is obvious to everyone, the explanation may appear a bit more subtle. The
footsteps reflect the past due to action of the second law of
thermodynamics: these footsteps can disappear without a cause due to
increasing entropy erasing information (solid blue lines in figure \ref{fig2}%
), but they cannot appear without a cause (i.e. without someone walking on
the beach), as this would contradict to the second law. Footsteps on a beach
can be washed out but cannot be "washed in". The scenario outlined by the
red dashed line in figure \ref{fig2} contradicts to the second law and is
impossible. Overall, the cause (a man walking) must precede the effect
(footsteps on the sand) due to constraints of the second law.

A photograph or something that we simply remember are subject to the same
constraints as the footsteps on a beach -- they are governed by the second
law and thus tell us about the past and not about the future. In the same
way, the other observed arrows of time must reflect the thermodynamic arrow:
surface waves cannot appear without being caused by a stone thrown into
water and electromagnetic waves cannot appear without a prior electric
impulse, but all these waves do not need any specific cause in the future
and can gradually dissipate forward in time due to the action of the second
law.

The example shown in figure \ref{fig2} illustrates that the second law
implies relative freedom in setting initial conditions by, say, walking on
the beach (although this freedom is not absolute) as compared to much lesser
freedom for achieving specific final conditions. The second law also
provides a transparent explanation for the practical version of the
causality principle. Consider two initial states A and A' of a thermodynamic
system at $t=t_{1}$ that result in essentially the same equilibrium state B
at $t=t_{2}$ (the phase volumes corresponding to these states are
schematically shown in figure \ref{fig3}). The evolutions A$\rightarrow $B
and A$^{\prime }$ $\rightarrow $B can be simulated by setting the initial
conditions at $t=t_{1}$ but not by setting the final conditions at $t=t_{2}$%
. Action of the second law of thermodynamics, which erases plenty of
macroscopically significant information, commonly makes setting the final
conditions ill-posed. Our preference for setting initial (and not final)
conditions, which is innately interpreted as the past causing the future, is
thus a natural outcome of the second law.

\section{Why does entropy increase?}

While the Boltzmann time hypothesis declares physical equivalence of the
causal and thermodynamic arrows of time, the mechanism behind the action of
the second law remains elusive. A number of factors (a few of these were
originally considered by Boltzmann \cite{K-PS2012T}) are commonly expected
to be relevant to this mechanism. These factors are discussed in this
section.

\subsection{Fixing the initial conditions}

The most common approach for demonstrating that entropy tends to increase in
time is enforcing specific initial conditions and leaving the final
conditions unconstrained. This may be perfectly satisfactory from the
perspective of statistical physics, but does not achieve our goal since
setting initial (as opposite to final) conditions presumes causality and
implicitly discriminates the directions of time. One cannot explain
causality in line with the Boltzmann time hypothesis by entropy increases
and, at the same time, deduce the propensity of the entropy to increase in
time from causality. The directions of time are often discriminated by very
strong intuitive perception of the "time flow" based on causality, which is
common among all people including scientists.

\subsection{The beginning and the end of the Universe}

The second law of the thermodynamics may be related to the expected temporal
boundary conditions imposed on the Universe: low entropy at the beginning
(i.e. uniform Big Bang) and, presumably, high entropy at the end \cite%
{PenroseBook}. While such boundary conditions should undoubtedly affect
global properties of entropy, the ubiquitous action of the second law raises
the question about the mechanism that brings the global boundary conditions
into every macroscopic process. Indeed, low initial and large final values
of entropy in the universe explain the overall entropy-increasing trend but,
on its own, cannot explain necessity of this increase in every experiment
where measures are taken to screen the experiment from the initial and final
conditions imposed on the universe. This, of course, does not preclude these
initial and final conditions to act through some mechanisms that are not
precisely known at present. Note that, in the present consideration, we must
avoid the trap of a priori discrimination of the directions of time by
presuming causality and setting only the low-entropy initial conditions.

Consider the example shown in figure \ref{fig4}: a system, which includes a
Carnot engine with two heat reservoirs --- the heater and the cooler --- and
a work reservoir \cite{Beretta1991}, is held in a remote location of the
universe. The engine preforms only two cycles (first, a cooling cycle and
then the corresponding work-producing cycle) and is in a dormant state for
thousand years before and after the cycles. In the dormant state, the
reservoirs and the engine working gas are fully isolated from each other and
are at their internal equilibria (or the there is a full equilibrium and
temperature difference between the hot and cold reservoirs is created by the
cooling cycle). Working of ideal Carnot cycles is reversible. This means
that if the engine performs a cooling cycle forward in time, it is seen
exactly as a work-producing cycle backward in time and vice versa. Observing
these ideal cycles does not reveal the direction of time --- even with the
most detailed specification of the experimental parameters, we cannot tell
if we are observing the cycles forward in time or backward in time.

The same experiment, however, would easily detect the action of the second
law if conducted with more realistic cycles: both the cooling cycle and the
work-producing cycle always underperform forward in time and overperform
backward in time. If the laws of physics are completely time-symmetric, it
must be the initial and final conditions imposed on the universe that
interfere with the operation of our Carnot engine. Prior to and after the
sequence of two cycles, the system remains statistically steady for thousand
years and, during these periods, even the full record of all molecular
motions remains time-symmetric. How can the system still "remember" the
initial state imposed on the universe? There is no obvious answer. There
might be a special field that is omnipresent everywhere and interferes with
the system, connecting it to numerous ongoing time-directional processes in
the rest of the universe. Another possibility is that the matter, which was
created in the early universe and presently forms the parts of the engine,
still "remembers" the directions of time (that is, the properties of matter
are not fully time-symmetric due to special conditions in the early
universe). In any case, there must be a specific mechanism affecting
performance of the system in a time-directional manner and it should be
possible, at least in principle, to detect this mechanism in experiments.

\subsection{Stosszahlansatz}

In his famous H-theorem, Boltzmann \cite{Boltzmann-book} demonstrated that
the hypothesis of molecular chaos, called the Stosszahlansatz, is sufficient
to ensure entropy increase in gases. This hypothesis discriminates the
directions of time assuming that the properties of colliding particles (i.e.
molecules) are statistically independent before (but not after) the
collision. Abe \cite{Abe2009} introduced a more general version of the
Stosszahlansatz applicable to non-extensive thermodynamic systems. The
Stosszahlansatz may have wider implications of a more generic statement:
outputs following an event tend to correlate more than the inputs preceding
this event. For example, scratches and dents on two cars have common
patterns only after their collision and not before. Price \cite{PriceBook}
often uses term "innocence" to characterise independence of incoming inputs.
Despite its importance and plausibility, the Stosszahlansatz (both in its
original and more general forms) largely remains a hypothesis rather than a
proven statement. Physically, the Stosszahlansatz, which is strongly
time-directional, is likely to be related to time-directional properties of
mixing, which are discussed below.

\subsection{Ergodic mixing in classical systems}

From the classical perspective, thermodynamic systems can be seen as dynamic
Hamiltonian systems of a very large dimension. As shown in figure \ref{fig5}%
a, evolution of these systems is time-reversible, preserves volumes in the
phase space (and, consequently, entropy, which is a logarithm of the phase
volume), and is expected to uphold a special mixing property, which is often
called ergodic mixing \cite{Ergo-theory}. This property requires that
overwhelming majority of dynamic trajectories occupy the phase space with
uniform average density irrespective of the initial (or final) conditions
imposed on the trajectories, provided these conditions are separated from
the current moment by a sufficiently long time interval. Under these
conditions any tiny coarsening of the trajectories results in the initial
phase volume expanding to the whole phase volume available to the system
(figure \ref{fig5}b). This coarsening occurs forward in time resulting in
entropy increase in this direction. Backward in time, the same process
represents "thinning" of the trajectories, reducing the occupied phase
volume and entropy. The rate of the phase volume increase is specified by
the so called Kolmogorov-Sinai dynamic entropy, which is determined by
dynamic properties of the system and by the presence (but generally not the
magnitude and characteristic time) of the coarsening. While it is possible
to explain average entropy increase in this model without coarsening by
setting initial (and not final) low-entropy conditions, this explanation
implicitly invokes causality --- this, as previously noted, would result in
circular logic in the present context.

\subsection{Quantum mixing: decoherence and collapses}

While the major equations of quantum mechanics are time-reversible and
entropy-preserving, irreversible quantum effects are associated with quantum
decoherence and collapses (see illustration in figure \ref{fig6} and the
Appendix). Since understanding of decoherence varies between publications,
this term is understood here as conversion of pure quantum states into mixed
quantum states irrespective of the underlying physical mechanisms. The
details are discussed in the Appendix. Introduction of decoherence replaces
reversible unitary evolutions of the Schr\"{o}dinger equation by the
irreversible evolutions of the Pauli master equation \cite%
{Pauli1928,Zeh2007,SciRep2016}. While conventional unitary evolutions of
quantum systems preserve entropy, decoherence increases entropy and the
maximal possible entropy is achieved in the maximally mixed state. Backward
in time, decoherence is interpreted as recoherence, which removes
uncertainty from the system and reduces entropy. As in the classical case,
the characteristic time of decoherence $\tau _{d}$ does not affect the rate
of entropy change (assuming that $\tau _{d}$ stays within its prescribed
physical range). This rate is determined by the Hamiltonian specifying
unitary interactions between major energy eigenstates. Nevertheless,
seemingly minor alteration of the time symmetry by decoherence has a
dramatic effect on the universe, introducing the arrow of time in kinetic
equations \cite{SciRep2016, Ent2017}.

Decoherence may involve both spontaneous (intrinsic) and environmental
mechanisms. The recent theories of decoherence tend to focus on decoherence
induced by unitary interactions with the environment \cite%
{Zurek2003,CT-P2009,Yukalov2012}. While environmental mechanisms can be
expected to play a significant role in decoherence, the conceptual treatment
of these interactions is based on unitary models, which cannot serve as
explanations for the time flow. Indeed, any theories that are based on
time-symmetric quantum mechanics rely on setting special initial conditions
for introducing temporal directionality, and this implicitly invokes
causality. These theories of environmental decoherence are useful in many
respects but cannot explain the arrow of time. At the same time, many
theories of spontaneous decoherence \cite{QTreview,Diosi2005,Beretta2015},
which introduce time-directional modifications of the unitary quantum
mechanics and, in principle, may explain the arrow of time, often lack
physical interpretation and experimental validation.

In terms of the governing equations, decoherence and collapse replace the
von Neumann equation for density matrices by the Lindblad equation (see the
Appendix). The former preserves the entropy, while the latter changes
entropy but preserves the trace and positiveness of the density matrix \cite%
{Zeh2007}. The Lindblad equation is more a framework for analysis rather
than a specific model. However, many specific models of decoherence and
collapse have been built around the Lindblad equation and a few of these
models introduce stochastic interference with quantum systems (e.g. Refs. 
\cite{Diosi1989,Zeh2007,Diosi2005,Beretta2015,Abe2017}). Most of these
models are phenomenological: i.e. they may predict how decoherence acts but
cannot explain why.

\section{The time primer}

The Boltzmann time hypothesis establishes a fundamental link between the
time arrow and entropy increase but, in its original form, does not
represent an experimentally testable theory, since the second law of
thermodynamics remains largely empirical and derived from engineering
practice and understanding. The missing physical explanation for the
directionality of the second law can be sought in statistical physics but,
as discussed in the previous section, this approach has to face significant
obstacles. Statistical physics incorporates classical and quantum mechanics
but, ultimately, has to deal with the original problem --- describing the
time-asymetric real world while using time-symmetric mechanical laws. The
existing approaches to justifying the second law of thermodynamics by
presuming causality can only lead us to circular arguments.

At this point we assume existence of a "time primer" --- a physical
mechanism (or mechanisms) that enacts irreversible processes associated with
the arrow of time (such as entropy increase, practical causality,
decoherence, collapse, etc. ). In simple terms we do not state that the
observed "time flow" is induced by the second law but rather that both the
"time flow" and the second law have a common physical cause, which is not
exactly known to modern science. The time primer may involve environmental
and spontaneous (intrinsic) mechanisms, interactions, collapses and
influences of the global boundary conditions --- we do not restrict its
mechanisms but expect that the physical action of the time primer can be
tested experimentally, at least in principle. It seems that the most
immediate measurable effect of the time primer should be quantum decoherence
(i.e. the time primer must at least provide a reasonable physical
explanation for quantum decoherence without presuming causality). In context
of environmental mechanisms, only those environmental interactions that must
be present intrinsically for enacting the direction of time are related to
the time primer. As the direction of time seems to be universal in the world
we know, the time primer is expected to be intrinsic or, at least,
intrinsically present.

The existence of the time primer is a logical extension of the Boltzmann
time hypothesis allowing us to avoid circular arguments, which are
implicitly based on causality and were occasionally used by Boltzmann to
explain the action of the second law. These circular arguments have only
become more common in recent times \cite{PriceBook}. The time primer
presumes the primacy of testable physical laws over causality, which is seen
more as a practical consequence of these laws than a priori universal
principle. Without specification of a particular mechanism, the time primer
is nothing more than a place holder for a physical theory of thermodynamic
time. Its role is in breaking circular arguments and becoming a "known
unknown" rather than an "unknown unknown". The time primer allows us to
discuss physical mechanisms enacting thermodynamic time without subscribing
for a particular model of decoherence or any other related physical process.
The possible mechanisms enacting the time primer may range from temporal
asymmetries hidden inside baryons to the action of gravity.

\subsection{Properties of the time primer}

While the time primer is an assumption that the mechanism enacting the
second law of thermodynamics and the perceived flow of time can be explained
by a physical, experimentally testable theory (which remains unknown at
present), we still can restrict properties of the time primer by applying
constraints based on common knowledge.\ The time primer must be
time-directional but should have a microscopic process of a very small
magnitude, which is difficult (but ultimately possible) to detect, since the
key time-symmetric theories of physics (e.g. classical and quantum
mechanics, special and general relativity) produce reasonably accurate
descriptions of reality. The time primer may reflect inherent properties of
matter or be a result of spontaneous breaking of time symmetry. A theory
explaining the time primer should not use causality as one of its postulates
and discriminate the directions of time by merely setting the initial (and
not final) conditions since, as previously noted, this would lead us to
circular reasoning.

Generally, we should expect time priming to involve both environmental and
spontaneous mechanisms, at least because distinguishing these mechanisms may
not always be possible. Consider a thermodynamic system, which is located in
a remote part of the universe in a way that prevents common thermodynamic
interactions with the universe. Should thermodynamic time stop for this
system? There is no evidence that this would be the case -- the pace of
irreversibly running time does not seem to depend on surroundings (for
example, the rate of nuclear decays does not seem to slow down when
environmental interference is reduced). Hence, if the universe still
influences the system, enforcing running thermodynamic time, this should be
some unavoidable interference, not constrained by locality and known
interaction mechanisms. Philosophically, this type of unavoidable
interference, which involves properties of the universe as the whole, is not
much different from intrinsic behaviour associated with the system. We call
this interference intrinsically present. After all, extending the concept of
environmental interference from specific measurable interactions with
devices and immediate environment to abstract remote interferences with the
whole of the universe removes the line of separation between environmental
and spontaneous mechanisms.

\subsection{Invariance of the time primer}

Since the time primer is expected to be a microscopic process violating time
symmetry, the invariant properties of this process represent a key question.
Presuming that matter and antimatter are fundamentally similar leads to two
possibilities:

\begin{enumerate}
\item the time primer is CP-preserving and CPT-violating or

\item the time primer is CPT-preserving and CP-violating.
\end{enumerate}

The first possibility corresponds to the symmetric (CP-invariant) extension
of thermodynamics from matter to antimatter, while the second possibility
corresponds to the alternative antisymmetric (CPT-invariant) extension \cite%
{KM-Entropy2014,SciRep2016}. As discussed in the rest of this work, these
alternatives do have implications for our understanding of the properties of
the universe. Note that the invariant properties of the time primer (i.e. CP
or CPT) are not necessarily determined by the invariant (CP or CPT)
properties of the unitary Hamiltonians \cite{SciRep2016}.

\subsection{Diosi-Penrose theory}

In this subsection we consider a possible mechanism for time priming by
gravitational collapse of quantum superposition states suggested by Diosi
and Penrose \cite{Diosi1987,Penrose1996,Diosi2005,Penrose2014a}. The
physical substance of the theory, which pertains to interactions of gravity
and quantum mechanics and distinguishes the Diosi-Penrose model from other,
more phenomenological approaches, is best explained by Penrose \cite%
{Penrose1996,Penrose2014a}, who discusses the fundamental difficulties of
unifying quantum mechanics with general relativity. Quantum superpositions
of masses should lead to superpositions of space-time geometries, which
should redefine understanding of space and time accepted in general
relativity and, according to Penrose \cite{Penrose1996,Penrose2014a}, lead
to non-linear interactions of superimposed states making these states
unstable. If $\Delta E$ is the self-gravitational energy associated with the
superposition states, than $\tau \sim \hbar /\Delta E$ would be
characteristic life time of these states, which becomes very small for
massive objects. Hence, quantum superposition states are not observed for
macroscopic objects --- Schr\"{o}dinger's cat must be either dead or alive.
Despite the extremely small amplitude of self-gravity, there can be fine
physical experiments that can detect existence of this mechanism, at least
in principle \cite{Penrose2014a}.

While the Diosi-Penrose theory seems to be closer than any other model to
discovering the physical mechanism behind the directionality of time, a
number of questions remain unanswered. First, both general relativity and
quantum mechanics are time-symmetric theories and it is not clear which
effect can discriminate the directions of time. Would this be a spontaneous
breaking of the time symmetry? If this is the case, why does time run in the
same direction everywhere? Penrose's analysis \cite{Penrose1996,Penrose2014a}
implicitly invokes causality and does not answer these questions. Second, is
the envisaged unified theory of quantum gravity expected to preserve the CPT
invariance of relativistic quantum mechanics? A negative answer would
deliver a devastating blow to our current understanding of the quantum
world, which is inherently linked to Feynman's interpretation of
antiparticles as particle equivalents moving backward in time.

\section{Detecting time priming in experiments\label{SecE}}

\subsection{Decoherence experiments}

Intensive experimental investigation of decoherence started only recently
and has been reviewed by Schlosshauer \cite{Schlosshauer2007}.\ The
successful experiments can be divided into three groups: decoherence of
qubit superposition states in superconductors\ \cite{Dec3Exp2003},
multi-slit wave interference for macro-molecules (with masses on the order
of a thousand of atomic mass units) \cite{Dec2Exp2004} and decoherence of
photonic coherent states in a cavity \cite{Dec1Exp2008}. It seems that all
these experiments are dominated by environmental decoherence. The
macro-molecule interference picture improves with decreased gas pressures 
\cite{Dec2Exp2003}, clearly indicating interactions of the macro-molecules
and the gas. Decoherence times of 20 ns (later of several $\mu $s \cite%
{Schlosshauer2007}, while qubits can be stored in semiconductors for more
than 30mins \cite{Dec4Exp2013}) have been achieved in superconductors \cite%
{Dec3Exp2003}. The characteristic times of decoherence for photons in a
cavity reach tens of ms and seem to be inversely proportional to the number
of photons \cite{Dec1Exp2008}. This, however, is likely to reflect bosonic
interaction terms rather than any spontaneous behaviour --- the number of
photons in the experiment (tens) is far too small while spontaneous
decoherence can be expected to become detectable only in larger systems.

While known experiments are likely to be dominated by environmental
decoherence \cite{Schlosshauer2007}, this does not mean that spontaneous
decoherence is not important, since the time primer, which determines the
direction of time, may have a very small magnitude and produce only a small
contribution in the experiments. To detect spontaneous mechanisms of
decoherence, we need to reduce the influence of the environment to a minimum
and increase the size of the system. We also note that some environmental
interferences may be intrinsically present and thus practically
indistinguishable from spontaneous mechanisms. The conventional wisdom of
achieving sufficient system sizes for detecting spontaneous decoherence has
nevertheless been questioned in some publications \cite{SpontDec2014} under
assumptions of decoherence driven by a stochastic interference. Penrose \cite%
{Penrose2014a} describes experimental setup developed by Dirk Bouwmeeste
that may be able to detect spontaneous decoherence predicted by
Diosi-Penrose theory with the use of a photon reflection from a tiny mirror.
While experimental validation of a physical theory, such as the one
suggested by Diosi and Penrose, may be difficult but ultimately possible, a
more generic examination physical nature of intrinsic decoherence is more
problematic due to absence of any prior information about the possible
magnitude of the process. Generally, it is impossible to search for
something that has an unknown magnitude and mechanism, but there might be a
possibility of detecting an unknown time primer due to its expected
invariant properties.

\subsection{Detection of the environmental time priming in CP-violating
processes}

Consider a quantum system in an environment with the overall Hamiltonian $%
\mathbf{H}_{U}$ represented in the Hilbert space $\mathcal{H}_{U}=\mathcal{H}%
\mathbf{\otimes }\mathcal{H}_{E}$ by 
\begin{equation}
\mathbf{H}_{U}\mathbf{=}\underset{\sim 1}{\underbrace{\mathbf{H}_{s}\mathbf{%
\otimes I}_{E}}}+\underset{\sim \varepsilon }{\underbrace{\mathbf{H}_{w}%
\mathbf{\otimes I}_{E}}}\mathbf{+}\underset{\sim 1}{\underbrace{\mathbf{%
I\otimes H}_{E}}}\mathbf{+}\underset{\sim \varepsilon }{\underbrace{\mathbf{H%
}_{\text{int}}}}  \label{HHH}
\end{equation}%
where $\mathbf{I}$\ denote the corresponding unit operators, the subscript "$%
E$" refers to the environment, the term $\mathbf{H}_{\text{int}}$ reflects
interaction of the system and environment. The state of the system and the
environment in space $\mathcal{H}_{U}$ is specified by the orthogonal basis $%
\left\vert s\right\rangle \mathbf{\otimes }\left\vert \beta \right\rangle $
denoted simply by $\left\vert s\right\rangle \left\vert \beta \right\rangle $%
. Here, $s=1,...,n$ enumerates the states of the system\ and $\beta
=1,...,n_{E}$ represents a very large number of states in the environment.
In many decaying systems (such as mesons), the system Hamiltonian has two
components the larger $\mathbf{H}_{s},$ which is associated with the strong
force and dominates the Hamiltonian, and $\mathbf{H}_{w},$ which is
associated with the weak force and is responsible for deviations from strong
eigenstates causing decays. The states of the system $s=\{k,f\}$ are divided
into initial states $k$ and the final (or intermediate) states $f$. \ Note
that $\left\langle k\right\vert \mathbf{H}_{s}\left\vert f\right\rangle
=0=\left\langle f\right\vert \mathbf{H}_{s}\left\vert k\right\rangle $ but $%
\left\langle k\right\vert \mathbf{H}_{w}\left\vert f\right\rangle \neq 0\neq
\left\langle f\right\vert \mathbf{H}_{w}\left\vert k\right\rangle $.\ The
small parameter $\varepsilon \ll 1$ indicates that the weak and
environmental interaction terms are both small compared the strong force.
While analyses neglecting environmental interactions in (\ref{HHH}) are
common \cite{PDG2012}, we can argue that unavoidable environmental influence
must be present in these experiments. Indeed, particle decays are
fundamentally irreversible, while unitary evolutions of the isolated quantum
system are cyclic and eventually must return to original states. Hence time
priming must be intrinsically present in decays. As discussed in the
Appendix, irreversibility can be introduced through spontaneous or
environmental decoherence. The former involves violations of quantum
mechanics, while the latter requires consideration of interactions with the
environment.

The suggested method of detecting unavoidable environmental interactions is
based on the following key assumptions: the system under consideration is
CPT-preserving (and may or may not be CP-violating), while interactions with
the environment are CP-preserving. Specifically, $\mathbf{H}_{s}$ is both
CPT- and CP-preserving and $\mathbf{H}_{w}$ is CPT-preserving and can be
CP-violating. \ The interactions with environment are assumed to be
CP-preserving and time-assymetric. The assumption of CP-invariance of
interactions with the environment is natural since overwhelming majority of
the physical processes in the universe are CP-invariant. In combination with
the temporal asymmetry of interactions with the environment, this may
produce an impression of a CPT-violation, although this is not the case.
Indeed, our treatment does not replace matter by antimatter in the
environment (which is obviously impossible) and this makes charge
conjugation C incomplete. Assuming that the composition of decaying
particles is matter/antimatter symmetric, such as in $K-$mesons whose most
stable configurations are approximately given by 
\begin{equation}
\left\vert K_{S}\right\rangle \approx \frac{\left\vert K\right\rangle
+\left\vert \bar{K}\right\rangle }{\sqrt{2}},\ \ \ \left\vert
K_{L}\right\rangle \approx \frac{\left\vert K\right\rangle -\left\vert \bar{K%
}\right\rangle }{\sqrt{2}}
\end{equation}%
introducing time priming by spontaneous decoherence would violate the
presumed CPT-invariance of the system. Hence, only environmental mechanism
of decoherence can be consistent with the key assumptions listed above.
Here, the initial states $k$ take two values: the kaon $\left\vert
K\right\rangle $ and the antikaon $\left\vert \bar{K}\right\rangle .$

The problem can be solved by seeking an asymptotic solution for effective
reduced Hamiltonian $\mathbf{H}_{\text{eff}}$ that acts in the space $%
\left\vert k\right\rangle \left\vert \beta \right\rangle $ in form of the
series $\mathbf{H}_{\text{eff}}=\mathbf{H}_{0}\mathbf{+}\varepsilon \mathbf{H%
}_{1}+\varepsilon ^{2}\mathbf{H}_{2}+...,\ \ $where$\ \left\langle k^{\prime
}\right\vert \mathbf{H}_{0}\left\vert k^{\prime \prime }\right\rangle
=\left\langle k^{\prime }\right\vert \mathbf{H}_{s}\mathbf{\otimes I}%
_{E}\left\vert k^{\prime \prime }\right\rangle $ for $k^{\prime },k^{\prime
\prime }=K,\bar{K}$ while the other terms are evaluated with the
Weisskopf-Wigner approximation \cite{K-PhysA}. The possible CPT violation,
which is indicated by the parameter 
\begin{equation}
\Lambda (\beta )=\left\langle K\right\vert \left\langle \beta \right\vert 
\mathbf{H}_{\text{eff}}\left\vert K\right\rangle \left\vert \beta
\right\rangle -\left\langle \bar{K}\right\vert \left\langle \beta
\right\vert \mathbf{H}_{\text{eff}}\left\vert \bar{K}\right\rangle
\left\vert \beta \right\rangle
\end{equation}%
appears only at the second order $\sim \varepsilon ^{2}$ of the expansion:%
\begin{equation}
\Lambda (\beta )=-\sum_{f}\left( \left( \left\langle K\right\vert
\left\langle \beta \right\vert \mathbf{H}_{\text{int}}\left\vert
f\right\rangle \left\vert \beta \right\rangle -\underset{}{\left\langle
f\right\vert \left\langle \beta \right\vert \mathbf{H}_{\text{int}%
}\left\vert K\right\rangle \left\vert \beta \right\rangle }\right) \frac{%
\left\langle K\right\vert \mathbf{H}_{w}\left\vert f\right\rangle
-\left\langle f\right\vert \mathbf{H}_{w}\left\vert K\right\rangle }{%
E_{0}-E_{f}+i\delta }\right)  \label{L1}
\end{equation}%
The sign of $\delta \rightarrow 0$ is selected to obtain only decaying
exponents. Equation (\ref{L1}) requires that $\mathbf{H}_{w}$ is
CPT-preserving and $\mathbf{H}_{\text{int}}$ is CP-preserving. If both $%
\mathbf{H}_{w}$ and $\mathbf{H}_{\text{int}}$ are CP-preserving, then $%
\Lambda (\beta )=0$ and there is no apparent CPT violation. Most quantum
systems are CP-preserving, and for these systems, the invariant properties
of weak and environmental interactions match each other so that the time
priming by environmental interactions remains invisible. However, in the
special cases of CP-violating systems, the environmental influence becomes
visible as an apparent CPT violation $\Lambda (\beta )\neq 0$ \cite{K-PhysA}%
. This violation is only apparent and not real since the system is,
intrinsically, CPT-preserving. Hence, it should be possible (at least in
principle) to detect the unavoidable environmental mechanisms of time
priming in CP-violating decays. Recent experiments seem to indicate that
larger than expected CPT discrepancies are indeed present in CP-violating
meson decays \cite{PDG2012,K-PhysA,Barbar2016}.

\subsection{Testing invariant properties of the spontaneous time primer}

At this point we return to consideration of spontaneous time priming.
Detecting spontaneous time priming is always difficult. While this should be
possible on the basis of a testing a specific theory or a mechanism (such as
the theory by Diosi and Penrose), here we avoid specific assumptions about
the mechanism of the time primer and ask only one question: is spontaneous
time priming CP- or CPT-invariant? Since radiation must be decoherence
neutral \cite{Ent2017}, spontaneous time priming needs to be associated with
the most fundamental properties of matter and antimatter leaving us two
possibilities: the directions of time priming for matter and antimatter are
either the same or opposite. This gives rise to two alternative possible
thermodynamics: symmetric and antisymmetric \cite%
{KM-Entropy2014,SciRep2016,Ent2017}. This alternative can be tested
experimentally --- we only need to create a thermodynamic antisystem (i.e. a
thermodynamic system made of antimatter) and screen it from the overriding
thermodynamic influence of the environment (see Refs. \cite%
{KM-Entropy2014,SciRep2016,Ent2017} for details). If the antisymmetric
thermodynamics is proven to be real, this would indicate the spontaneous
mechanisms of time priming.

\section{Conclusions}

Universal causality, which for some time was presumed to be a priori of
rational thinking about the universe, leads to a spectrum of unresolvable
problems and has gradually been replaced by more practical interpretations
of causality that does not stay above the physical laws but is linked to
these laws or derived from them. In this context, Boltzmann's time
hypothesis identifying the observed direction of time with the second law of
thermodynamics seems most promising. We also expect that the action of the
second law and the arrow of time are initiated by physical process of a very
small scale that, at least in principle, can be detected in the experiments.
This physical process is termed here as the "time primer". While the action
of the time primer is extremely difficult to distinguish from other
interactions, the recent works on experimental investigation of decoherence
gradually progress in this direction. Besides these experiments, it seems
that the intrinsically present action of the time primer can be detected in
CP-violating (and CPT-preserving) decays of elementary particles and should
be seen as apparent CPT violations.

\bibliographystyle{unsrt}
\bibliography{Law3}

\appendix

\section{Appendix: decoherence and collapse}

The arrow of time is believed to be closely connected to irreversible
processes of quantum mechanics \cite{PenroseBook,Zeh2007}. While
interpretation of these processes is a difficult problem on its own, the
reader can be easily confused by inconsistencies of definitions used in
different publications. For example, decoherence is often defined only as an
environment-induced process, while spontaneous behaviour is related
exclusively to quantum collapses \cite{PhysPhil2009}. Other publications
(e.g. \cite{Stamp2012}), however, refer to both environmental and intrinsic
decoherences. This appendix clarifies the use of the terms "decoherence" and
"collapse" in the present context of considering the Boltzmann time
hypothesis.

Restricting term "decoherence" only to environmental processes would be
rather inconvenient, since both types of decoherence, environmental and
spontaneous, produce very similar effects. In addition, both types of
decoherence are likely to act in combination with each other and, as
discussed in Section \ref{SecE}, are very difficult to distinguish
experimentally. Decoherence is understood here is a process of converting
pure quantum states into mixed states or increasing the degree of mixing in
already mixed states, irrespective of the physical mechanism(s) responsible
for the process. Decoherence increases the entropy $S$ of the system.
Collapse is reduction of the wave function to a particular state during
measurements or interactions with macroscopic objects and, besides
decoherence, may involve other effects. If spontaneous decoherence is to be
exclusively associated with collapse, the latter term needs further
qualification, as discussed below. While the author of the present work
tends to use the terms "spontaneous decoherence" and "intrinsic decoherence"
synonymously, there is a semantic difference between these terms that is
worth noting. "Spontaneous" means "without an external cause" while
"intrinsic" implies impossibility to avoid something by changing external
conditions. Hence, interactions of a quantum system with an omnipresent
time-priming field (assuming that such a field exists) are intrinsic but not
spontaneous.

\subsection{Modelling decoherence}

A general framework for consideration of models deviating from unitary
evolution is given by the Lindblad equation 
\begin{equation}
\underset{\text{von Neumann Eq.}}{\underbrace{i\hbar \frac{\partial \mathbf{%
\rho }}{\partial t}=\left[ \mathbf{H},\mathbf{\rho }\right] }}-\underset{%
\text{Lindblad operator}}{\underbrace{\frac{i}{2}\sum_{j=1}^{n^{2}-1}\left( 
\mathbf{L}_{j}^{\dag }\mathbf{L}_{j}\mathbf{\rho +\rho L}_{j}^{\dag }\mathbf{%
L}_{j}-2\mathbf{L}_{j}\mathbf{\rho L}_{j}^{\dag }\right) }}  \label{Lind}
\end{equation}%
for density matrices $\mathbf{\rho }$ that are expressed by the sum of the
outer products of the wave functions $\psi _{k}$%
\begin{equation}
\mathbf{\rho }=\sum_{k=1}^{N}p_{k}\left\vert \psi _{k}\right\rangle
\left\langle \psi _{k}\right\vert  \label{rho}
\end{equation}%
The state specified by this equation is mixed if $N>1$ and pure if $N=1$.
The mixture is maximally ranked if $N$ reaches the dimension of the system $%
N=n.$ The state is maximally mixed if $N=n$ and $p_{k}=1/n$ for $k=1,2,...,n$%
. The maximally mixed state corresponds to microcanonical thermodynamic
equilibrium in macroscopic objects, attaining the maximal value of the
entropy $S,$ which is defined in quantum mechanics as%
\begin{equation}
S=-k_{B}\func{Tr}\left( \mathbf{\rho }\ln \mathbf{(\rho )}\right) ,
\label{SS}
\end{equation}%
where $k_{B}$ is the Boltzmann constant. The von Neumann equation for $%
\mathbf{\rho }$ corresponds to the Schr\"{o}dinger equation for the wave
functions $\psi _{k}$, which represents reversible, unitary evolution and
preserves the entropy: $dS/dt=0$. The Lindblad operator in equation (\ref%
{Lind}) introduces changes of entropy $dS/dt\neq 0$ but preserves the trace $%
\func{Tr}\left( \mathbf{\rho }\right) =1$ and positiveness of the density
matrix $\mathbf{\rho }$. While the general form of equation (\ref{Lind})
does not guarantee monotonic increase of the entropy, there would be many
specific forms of this equation that ensure that $dS/dt\geq 0$ (e.g. \cite%
{Beretta2015,Abe2017}).

The Lindblad equation can specify various non-unitary behavious including
decoherence and collapse. Decoherence corresponds to disappearance of
non-diagonal elements in the density matrix $\mathbf{\rho }$: 
\begin{equation}
\left[ 
\begin{array}{ccc}
\rho _{11} & \cdots & \rho _{n1} \\ 
\vdots & \ddots & \vdots \\ 
\rho _{1n} & \cdots & \rho _{nn}%
\end{array}%
\right] \longrightarrow \left[ 
\begin{array}{ccc}
\rho _{11} &  & 0 \\ 
& \ddots &  \\ 
0 &  & \rho _{nn}%
\end{array}%
\right]
\end{equation}%
If a system interacts with environment (whose Hilbert space is denoted by $%
\mathcal{H}_{E})$ so that the "universe" is characterised by a Hilbert space 
$\mathcal{H}_{U}=\mathcal{H}\otimes \mathcal{H}_{E}$, then the density
matrix of the system $\mathbf{\rho }$ can be obtained from the density
matrix of the universe $\mathbf{\rho }_{U}$ by tracing out the degrees of
freedom associated with the environment 
\begin{equation}
\mathbf{\rho =}\func{Tr}_{E}\left( \mathbf{\rho }_{U}\right)
\end{equation}%
As it has been shown by Zurek \cite{Zurek1982}, very weak interactions of
the system with environment are sufficient to cause prompt decoherence of
the system even if the process is controlled by the conventional Schr\"{o}%
dinger equation. There are no doubts that environmental interpretation of
the decoherence is attractive --- the complicated problem of modifying
unitary dynamics in equation (\ref{Lind}) is apparently avoided. The
environmental approach to decoherence has been advocated by many \cite%
{Joos2003,Schlosshauer2007,CT-P2009,Yukalov2012} and quite rightfully so. It
is most likely that environmental decoherence dominates in many practical
situations (see section \ref{SecE}). We, however, cannot forget that
discarding intrinsic decoherence completely would lead our analysis into
unresolvable problems. Indeed, presuming absence of any directional
time-priming mechanism must be compensated by the de-facto introduction of
causality and, as previously noted, this leads us to circular arguments.
Zurek's \cite{Zurek1982} decoherence theory is necessarily based on
causality. Another problem is that the entropy increase --- a fundamental
process that, according to Boltzmann's time hypothesis, has a determining
influence on temporal properties of the universe --- does not occur in the
universe according to this model (any entropy increase in the system is
compensated by exactly the same entropy decrease in the environment).
Arguments based on treating the environment as an infinite sequence of
nested environments (akin Russian dolls) may hide the entropy problem for
the system but does not form a realistic model reflecting the overall
entropy increase in the universe.

\subsection{ Possible stages of collapse}

Quantum measurements and, in fact, many other forms interactions of quantum
systems with macroscopic objects (such as Schr\"{o}dinger's cat) lead to
collapse of the wave function \cite{PenroseBook}. As discussed below, this
collapse is related to but is not fully synonymous to decoherence. In our
analysis, we intend to save the cat's life and treat quantum/classic
interactions as a measurement problem. We broadly follow von Neumann's
theory of quantum measurement \cite{VonNeumann1955}, but consider variations
around the Copenhagen interpretation associated with decoherence.

\begin{description}
\item[0. Initial state.] Initially, the quantum system is in a superposition
state and the measurement apparatus is in the "ready" state $\left\vert
a_{0}\right\rangle $. The corresponding wave functions of the system $\psi
_{s}=c_{\uparrow }\left\vert \uparrow \right\rangle +c_{\downarrow
}\left\vert \downarrow \right\rangle $ and the apparatus $\psi
_{a}=\left\vert a_{0}\right\rangle $ indicate that the initial combined
system-apparatus state is given by%
\begin{equation}
\Psi _{0}=\psi _{s}\otimes \psi _{a}=c_{\uparrow }\left\vert \uparrow
\right\rangle \left\vert a_{0}\right\rangle +c_{\downarrow }\left\vert
\downarrow \right\rangle \left\vert a_{0}\right\rangle  \label{PSI0}
\end{equation}%
Here, we assume for simplicity that the quantum system has only two states $%
\left\vert \uparrow \right\rangle $ and $\left\vert \downarrow \right\rangle
,$ but avoid the trivial cases of $c_{\uparrow }=0$ and $c_{\downarrow }=0$.
All states are presumed orthogonal. The corresponding density matrix $%
\mathbf{\rho }_{0}\mathbf{=}\left\vert \Psi _{0}\right\rangle \left\langle
\Psi _{0}\right\vert $ is given by%
\begin{equation}
\mathbf{\rho }_{0}=c_{\uparrow }^{\ast }c_{\uparrow }\left\vert \uparrow
\right\rangle \left\vert a_{0}\right\rangle \left\langle \uparrow
\right\vert \left\langle a_{0}\right\vert +c_{\downarrow }^{\ast
}c_{\downarrow }\left\vert \downarrow \right\rangle \left\vert
a_{0}\right\rangle \left\langle \downarrow \right\vert \left\langle
a_{0}\right\vert +c_{\downarrow }^{\ast }c_{\uparrow }\left\vert \uparrow
\right\rangle \left\vert a_{0}\right\rangle \left\langle \downarrow
\right\vert \left\langle a_{0}\right\vert +c_{\uparrow }^{\ast
}c_{\downarrow }\left\vert \downarrow \right\rangle \left\vert
a_{0}\right\rangle \left\langle \uparrow \right\vert \left\langle
a_{0}\right\vert
\end{equation}%
where and the asterisk "$\ast $" denotes the complex conjugates. The initial
superposition state corresponds to zero entropy 
\begin{equation}
S_{0}=-k_{B}\func{Tr}\left( \mathbf{\rho }_{0}\ln \mathbf{(\rho }_{0}\mathbf{%
)}\right) =0
\end{equation}

\item[1. Unitary co-evolution.] Interactions between the system and the
apparatus result in a transformation $\Psi _{0}\longrightarrow \Psi _{1}$%
\begin{equation}
\Psi _{1}=c_{\uparrow }\left\vert \uparrow \right\rangle \left\vert
a_{\uparrow }\right\rangle +c_{\downarrow }\left\vert \downarrow
\right\rangle \left\vert a_{\downarrow }\right\rangle  \label{PSI1}
\end{equation}%
implying that the corresponding density matrix $\mathbf{\rho }_{1}\mathbf{=}%
\left\vert \Psi _{1}\right\rangle \left\langle \Psi _{1}\right\vert $ is%
\begin{equation}
\mathbf{\rho }_{1}=c_{\uparrow }^{\ast }c_{\uparrow }\left\vert \uparrow
\right\rangle \left\vert a_{\uparrow }\right\rangle \left\langle \uparrow
\right\vert \left\langle a_{\uparrow }\right\vert +c_{\downarrow }^{\ast
}c_{\downarrow }\left\vert \downarrow \right\rangle \left\vert a_{\downarrow
}\right\rangle \left\langle \downarrow \right\vert \left\langle
a_{\downarrow }\right\vert +c_{\downarrow }^{\ast }c_{\uparrow }\left\vert
\uparrow \right\rangle \left\vert a_{\uparrow }\right\rangle \left\langle
\downarrow \right\vert \left\langle a_{\downarrow }\right\vert +c_{\uparrow
}^{\ast }c_{\downarrow }\left\vert \downarrow \right\rangle \left\vert
a_{\downarrow }\right\rangle \left\langle \uparrow \right\vert \left\langle
a_{\uparrow }\right\vert
\end{equation}%
where the sates of apparatus $\left\vert a_{\uparrow }\right\rangle $ and $%
\left\vert a_{\downarrow }\right\rangle $ are interpreted as measuring $%
\left\vert \uparrow \right\rangle $ and measuring $\left\vert \downarrow
\right\rangle $. This transformation is conventionally assumed unitary (more
precisely, it can be made unitary by proper selection of responses to other
initial states, say $\left\vert \uparrow \right\rangle \left\vert
a_{\uparrow }\right\rangle $ or $\left\vert \uparrow \right\rangle
\left\vert a_{\downarrow }\right\rangle $ that are not of interest here).
The unitary transformation can be more complicated than specified by (\ref%
{PSI1}), as long as $\left\vert a_{\uparrow }\right\rangle \neq \left\vert
a_{\downarrow }\right\rangle $ and the initial states of the system can be
distinguished. Unitary evolutions do not change the value of the entropy 
\begin{equation}
S_{1}=-k_{B}\func{Tr}\left( \mathbf{\rho }_{1}\ln \mathbf{(\rho }_{1}\mathbf{%
)}\right) =0
\end{equation}

\item[2. Decoherence.] Interaction with the apparatus, which is a
macroscopic object, involves another process $\Psi _{1}\longrightarrow \Psi
_{2}$, which is conventionally called decoherence. This process is
non-unitary and/or can involve environmental interference. According to
Zurek's theory of decoherence \cite{Zurek1982,Zurek2003} even the smallest
interactions of quantum systems with macroscopic objects that do not
transfer any substantial energy can result in a loss of coherence, that is 
\begin{equation}
\Psi _{2}=\Theta _{\uparrow }c_{\uparrow }\left\vert \uparrow \right\rangle
\left\vert a_{\uparrow }\right\rangle +\Theta _{\downarrow }c_{\downarrow
}\left\vert \downarrow \right\rangle \left\vert a_{\downarrow }\right\rangle
\label{PSI2}
\end{equation}%
where $\left\vert \Theta _{k}\right\vert =1$, $k=\{\uparrow ,\downarrow \}$
and the phases $\theta _{k}=\ln (\Theta _{k})/i$ become unpredictable or,
practically, random due to decohering interference of the system with
macroscopic objects. As the result of decoherence, the density matrix
becomes diagonal $\mathbf{\rho }_{1}\longrightarrow \mathbf{\rho }_{2}$
where 
\begin{equation}
\mathbf{\rho }_{2}=\overline{\left\vert \Psi _{2}\right\rangle \left\langle
\Psi _{2}\right\vert }=c_{\uparrow }^{\ast }c_{\uparrow }\left\vert \uparrow
\right\rangle \left\vert a_{\uparrow }\right\rangle \left\langle \uparrow
\right\vert \left\langle a_{\uparrow }\right\vert +c_{\downarrow }^{\ast
}c_{\downarrow }\left\vert \downarrow \right\rangle \left\vert a_{\downarrow
}\right\rangle \left\langle \downarrow \right\vert \left\langle
a_{\downarrow }\right\vert
\end{equation}%
and the off-diagonal terms disappear (or, depending on the interpretation,
effectively disappear or disappear on average) due to the absence of the
phase correlations $\overline{\Theta _{\uparrow }^{\ast }\Theta _{\downarrow
}}=0,$ where the overbar denotes averaging over the random values. The
corresponding entropy increases $S_{1}\longrightarrow S_{2},$ where%
\begin{equation}
S_{2}=-k_{B}\func{Tr}\left( \mathbf{\rho }_{2}\ln \mathbf{(\rho }_{2}\mathbf{%
)}\right) =-k_{B}\left( p_{\uparrow }\ln p_{\uparrow }+p_{\downarrow }\ln
p_{\downarrow }\right) >0,
\end{equation}%
and $p_{\uparrow }=c_{\uparrow }^{\ast }c_{\uparrow },$ $\ p_{\downarrow
}=c_{\downarrow }^{\ast }c_{\downarrow }$ are non-negative real numbers,
satisfying $p_{\uparrow }+p_{\downarrow }=1$ due to normalisation of the
wave function.

Our consideration needs some clarifications at this stage. Realistically,
the apparatus is a macroscopic object. That is, in addition to\ the
measurement states $\left\vert a_{k}\right\rangle =\{\left\vert
a_{0}\right\rangle ,$ $\left\vert a_{\uparrow }\right\rangle ,$ $\left\vert
a_{\downarrow }\right\rangle ,...\}$, the apparatus is characterised by a
very large number $n$ of microscopic states $\left\vert q_{j}\right\rangle ,$
$j=1,...,n$ so that the apparatus has, at least $3n$ quantum states $%
\left\vert a_{k}\right\rangle \left\vert q_{j}\right\rangle .$ Since the
apparatus is macroscopic, its state cannot significantly deviate from the
maximally mixed state, which represents a thermodynamic equilibrium under
microcanonical conditions (for example, as a macroscopic object, the
apparatus cannot be in a superposition state, say, $\psi _{a}=\Sigma
_{j}q_{j}\left\vert a_{0}\right\rangle \left\vert q_{j}\right\rangle $ due
to almost instant action of the time primer). Even very weak interactions of
the quantum system with a macroscopic object can lead to very fast
decoherence, whose joint effect with unitary evolution can be characterised
by the transition $\mathbf{\rho }_{0}\longrightarrow \mathbf{\rho }_{2}.$
The decoherence of the quantum system states is not spontaneous --- it is
induced by the measuring apparatus. As a macroscopic object, the apparatus
is influenced by the time primer to stay close to its maximally mixed state.
The question of whether this time primer is intrinsic to the apparatus or
induced by the environment surrounding the apparatus (or both) does not
affect our consideration.

\item[3. Latent collapse.] The next presumed stage in quantum measurement is
collapse of the wave function $\Psi _{2}\longrightarrow \Psi _{3}$ with the
probability $P$ specified by the Born rule 
\begin{equation}
\Psi _{3}=\left\{ 
\begin{array}{cc}
\left\vert \uparrow \right\rangle \left\vert a_{\uparrow }\right\rangle , & 
P=c_{\uparrow }^{\ast }c_{\uparrow }=p_{\uparrow } \\ 
\left\vert \downarrow \right\rangle \left\vert a_{\downarrow }\right\rangle ,
& P=c_{\downarrow }^{\ast }c_{\downarrow }=p_{\downarrow }%
\end{array}%
\right.
\end{equation}%
We must emphasise that, at this stage, we do not know (and, as discussed
below, cannot possibly know) the outcome of the measurement (i.e. $%
\left\vert \uparrow \right\rangle \left\vert a_{\uparrow }\right\rangle $ or 
$\left\vert \downarrow \right\rangle \left\vert a_{\downarrow }\right\rangle 
$), although the wave function $\Psi $ has already collapsed into one of the
states $\left\vert \uparrow \right\rangle \left\vert a_{\uparrow
}\right\rangle $ or $\left\vert \downarrow \right\rangle \left\vert
a_{\downarrow }\right\rangle $. The density matrix is then represented by
the densities of two states $\mathbf{\rho }_{\uparrow }=\left\vert \uparrow
\right\rangle \left\vert a_{\uparrow }\right\rangle \left\langle \uparrow
\right\vert \left\langle a_{\uparrow }\right\vert $ and $\mathbf{\rho }%
_{\downarrow }=\left\vert a_{\downarrow }\right\rangle \left\langle
\downarrow \right\vert \left\langle a_{\downarrow }\right\vert $ taken with
the corresponding probabilities $p_{\uparrow }$ and $p_{\downarrow }$: 
\begin{equation}
\mathbf{\rho }_{3}=p_{\uparrow }\mathbf{\rho }_{\uparrow }+p_{\downarrow }%
\mathbf{\rho }_{\downarrow }=p_{\uparrow }\left\vert \uparrow \right\rangle
\left\vert a_{\uparrow }\right\rangle \left\langle \uparrow \right\vert
\left\langle a_{\uparrow }\right\vert +p_{\downarrow }\left\vert \downarrow
\right\rangle \left\vert a_{\downarrow }\right\rangle \left\langle
\downarrow \right\vert \left\langle a_{\downarrow }\right\vert
\end{equation}%
Note that $\mathbf{\rho }_{3}=\mathbf{\rho }_{2}$ and 
\begin{equation}
S_{3}=-k_{B}\func{Tr}\left( \mathbf{\rho }_{3}\ln \mathbf{(\rho }_{3}\mathbf{%
)}\right) =S_{2}>0
\end{equation}%
The reader may notice that there is only a fine logical line separating the
states 2 and 3. Practically, the outcomes of decoherence and latent collapse
are very similar but, still, there is a difference. Indeed, in the state 2,
the uncertainty between states $\left\vert \uparrow \right\rangle $ and $%
\left\vert \downarrow \right\rangle $ is deeply embedded into physics of the
system, while in the state 3, the system is already in one of the states $%
\left\vert \uparrow \right\rangle $ or $\left\vert \downarrow \right\rangle $%
, we simply do not know which one is that. We can call randomness and
associated probability \textit{ontic }or\textit{\ aleatory} (i.e. related to
objective physical uncertainty) in the first case, and \textit{epistemic}
(i.e. related to lack of knowledge) in the second case. While the existence
of intuitive and philosophical differences between ontic and epistemic
interpretations of probability needs to be acknowledged, quantum mechanics
tends to treat states 2 and 3 as physically equivalent (or, at least,
equivalent for all practical purposes), and has good reasons for this.
Indeed, no measurement of any observable physical quantity $A$ can possibly
distinguish these two states since $\mathbf{\rho }_{3}=\mathbf{\rho }_{2}$
and $A=\func{Tr}\left( \mathbf{\rho }_{2}\mathbf{A}\right) =\func{Tr}\left( 
\mathbf{\rho }_{3}\mathbf{A}\right) $. The equivalence of ontic and
epistemic probabilities assigns objective interpretation to epistemic
probability in quantum measurements --- the observer cannot obtain knowledge
without interfering with the system. This leads us to conclusion that
epistemic uncertainty is objectively present in the surrounding world since,
under conditions of non-interference, it cannot be distinguished from
"genuinely objective" ontic uncertainty. This approach is reasonable in
application to quantum measurements but may cause unease for many people if
it is interpreted as a general principle applicable to macroscopic world.
Einstein's famous "I like to think the moon is there even if I am not
looking at it" insists that physical reality is one thing, while knowing or
not about this reality is another.

\item[4. Observable collapse.] This is the final stage of the measurement
process, when the result of the measurement becomes known (or can become
known) due to producing a macroscopic effect.\ This stage can lead to one of
the two possible outcomes: 
\begin{equation}
4^{\prime })\text{\ \ }\Psi _{4^{\prime }}=\left\vert \uparrow \right\rangle
\left\vert a_{\uparrow }\right\rangle ,\ \ \mathbf{\rho }_{4^{\prime
}}=\left\vert \uparrow \right\rangle \left\vert a_{\uparrow }\right\rangle
\left\langle \uparrow \right\vert \left\langle a_{\uparrow }\right\vert
\end{equation}%
or 
\begin{equation}
4^{\prime \prime })\text{\ \ }\Psi _{4^{\prime \prime }}=\left\vert
\downarrow \right\rangle \left\vert a_{\downarrow }\right\rangle ,\ \ 
\mathbf{\rho }_{4^{\prime \prime }}=\left\vert \downarrow \right\rangle
\left\vert a_{\downarrow }\right\rangle \left\langle \downarrow \right\vert
\left\langle a_{\downarrow }\right\vert
\end{equation}%
with the corresponding frequencies proportional to $p_{\uparrow }$ and $%
p_{\downarrow }$ in repeated experiments. The outcomes of the observed
collapse are very different from the outcomes of decoherence. In both cases $%
4^{\prime }$ and $4^{\prime \prime }$ the entropy $S$ is zero 
\begin{equation}
S_{4^{\prime }}=-k_{B}\func{Tr}\left( \mathbf{\rho }_{4^{\prime }}\ln 
\mathbf{(\rho }_{4^{\prime }}\mathbf{)}\right) =0=-k_{B}\func{Tr}\left( 
\mathbf{\rho }_{4^{\prime \prime }}\ln \mathbf{(\rho }_{4^{\prime \prime }}%
\mathbf{)}\right) =S_{4^{\prime \prime }}
\end{equation}%
This apparent reduction of entropy indicates that other factors must be
present in observing the results of measurement. We may consider
interference of an observer, which has two possible outcomes of the
measurement: $\left\vert \uparrow \right\rangle \left\vert a_{\uparrow
}\right\rangle \left\vert o_{\uparrow }\right\rangle $ or $\left\vert
\downarrow \right\rangle \left\vert a_{\downarrow }\right\rangle \left\vert
o_{\downarrow }\right\rangle ,$ where $\left\vert o_{\uparrow }\right\rangle 
$ and $\left\vert o_{\downarrow }\right\rangle $ correspond to "observing $%
\uparrow $" and "observing $\downarrow $". We will try, however, to keep not
only cats but also humans out of the experiments (as both are known to have
"bad" behaviours that are rather difficult to predict) and focus on the
measuring apparatus as a thermodynamic object.

Since, according to the second law, entropy cannot have a consistent
unmitigated decrease $0<S_{3}\longrightarrow S_{4}=0$ even by a small amount
(fluctuations are possible, of course, but they are inconsistent), there
must be an entropy increase $\Delta S_{a}$ in the apparatus to complete the
measurement. Hence, some energy $\Delta E_{1}=T_{a}\Delta S_{a},$ where $%
T_{a}$ is the temperature of the apparatus, must be transferred to the
apparatus in the course of the measurement. (For simplicity, the possibility
of cooling down some of the apparatus parts, which indeed can improve the
detection, is not considered). Since $\Delta S_{a}\geq S_{3}$, and $%
S_{3}\sim k_{B}$ (we can assume $p_{\uparrow }=p_{\downarrow }=1/2$\ for the
estimates), the transferred energy must be sufficiently large $\Delta
E_{1}\gtrsim k_{B}T_{a}$. Indeed, the measuring apparatus cannot detect any
quantity that has energy effect comparable to $k_{B}T_{a}/2$ --- the average
energy of thermal fluctuations associated with each excited degree of
freedom. Detection needs a substantial initial energy effect $\Delta
E_{1}\gg k_{B}T_{a}/2$ that stands above the background of thermal
fluctuations.

The initial effect needs to be amplified (say $\Delta E_{1}\longrightarrow
\Delta E_{2}\gg \Delta E_{1}$) to reach macroscopic level and become easily
detectable. A macroscopic object must stay reasonably close to a maximally
mixed state since, otherwise, this process would spontaneously reduce
entropy and violate the second law. As a thermodynamic object, the apparatus
has a large number of degrees of freedom. While most of these degrees are
microscopic, the final measurement state must be a macroscopic parameter or,
otherwise, the results of measurement would remain undetected. Macroscopic
parameters also fluctuate \cite{LL5} but these fluctuations are very small $%
\sim k_{B}T_{a}$ compared to the average energy effect of the measurement $%
\sim \Delta E_{2}.$ Any fluctuation or uncertainty that is much larger then $%
k_{B}T_{a}$\ per degree of freedom is highly improbable according to the
Gibbs distribution and, if created, would be promptly attenuated by the
thermodynamic interactions. Once the measured value is amplified to
macroscopic level, it becomes detectable with a high degree of certainty.
The presence of an actual observer is not important --- in this context
"observable" means reaching macroscopic scales and producing a macroscopic
effect which can be observed.
\end{description}

While decoherence is a very rapid process involving only very weak
interactions with almost no energy transferred \cite%
{Zurek1982,Zurek2003,Yukalov2012}, the results of measurements can become
known only after some time period, associated with energy transfer from the
system to the apparatus (since, as discussed above, immediate availability
of the results without energy transfer to the apparatus contradicts to the
second law of thermodynamics). Therefore, information associated with the
measurement is not available immediately after decoherence. Most acts of
decoherence (and associated latent collapse) are not related to any
measurements, never produce any macroscopic effect and are never observed.
These acts do not involve observable collapses but, instead, increase
uncertainty and entropy (i.e. $S_{0}\longrightarrow S_{3}$). Here,
decoherences and latent collapses are understood as fundamental microscopic
processes that underpin the macroscopic conduct of the second law. Since
entropy increases on average everywhere in the universe, latent collapses
must dominate observable collapses. This interpretation represents a
thermodynamics-centred perspective on decoherence and collapse dictated by
the Boltzmann time hypothesis.

\subsection{Everett's many-worlds interpretation}

It is useful to consider an alternative treatment of quantum measurements
associated with Everett's many-worlds interpretation \cite%
{Everett1957,Dewitt1970}. This interpretation replaces collapse by the
assumption of complete independence of the two states $\left\vert \uparrow
\right\rangle $ and $\left\vert \downarrow \right\rangle $ after their
decoherence. This independence persists forward in time creating two
different "worlds" that remain independent of each other. Due to
entanglement between the system, the apparatus, the observer, and, in fact,
the rest of the universe, these worlds are slightly different having $%
\left\vert \uparrow \right\rangle \left\vert a_{\uparrow }\right\rangle
\left\vert o_{\uparrow }\right\rangle $ in the world $\mathcal{W}_{\uparrow
} $ and $\left\vert \downarrow \right\rangle \left\vert a_{\downarrow
}\right\rangle \left\vert o_{\downarrow }\right\rangle $ in the world $%
\mathcal{W}_{\downarrow }$. As subjective observers, we can find ourselves
only in one of these worlds $\mathcal{W}_{\uparrow }$ or $\mathcal{W}%
_{\downarrow }$ with the corresponding probabilities $p_{\uparrow }$ and $%
p_{\downarrow }$. While intellectually stimulating and original, the
many-worlds interpretation is practically equivalent to collapse, as long as
we do not know and cannot possibly predict a priori which of the worlds, $%
\mathcal{W}_{\uparrow }$ or $\mathcal{W}_{\downarrow },$ we will find
ourselves in. Each of these branched worlds will split further and further
with every irreversible act of quantum decoherence occurring anywhere in the
universe.

Assigning a world split to every irreversible act of quantum decoherence
corresponds the strong version\ of the Everett interpretation. Since a
subjective observer finds himself only in one of the worlds, where only one
of the states $\left\vert \uparrow \right\rangle $ or $\left\vert \downarrow
\right\rangle $ is present, this decoherence thus appears in each of these
worlds as a collapse. The key role of the strong version is thus in
explaining the physical correspondence between decoherence and collapse.
Since most of the collapses are latent, most of the splits produce
macroscopically indistinguishable worlds. The subjective observer is located
in one of the worlds but cannot know which one as many of the worlds appear
to be identical under all possible observations. This excessive redundancy
of the world splits can be removed by introducing a weak version of the
Everett interpretation. This version proclaims that the world splits
correspond only to observable collapses and, according to this version, all
branched worlds are macroscopically different (the difference can be small
or large but it must be macroscopic and observable).

Whether we might have many worlds or just one world in reality, the Everett
interpretation is a very useful way of thinking about quantum mechanics
replacing mysterious quantum randomness by branching the worlds, which, at
least, is a useful way of thinking about randomness. This interpretation is
attractive because it saves deterministic causality, which is deeply
embedded into our intuition, from the harsh and unpredictable real world.
Indeed, the future of the collective multiverse $\{\mathcal{W}_{\uparrow }$, 
$\mathcal{W}_{\downarrow },$...\} is certain and fully determined by the
past. It is only our subjective "I" that does not know which of these worlds
it is going to find itself in. However, in the context of the present work,
intuitive causality plays a negative role. Hence, we need to look at the
Everett interpretation from a more time-symmetric perspective. The weak
version of the interpretation is implied in the rest of the section.

Consider the initial state of the system and apparatus before the
measurement 
\begin{equation}
\Psi _{0}^{\prime }=c_{\uparrow }\left\vert \uparrow \right\rangle
\left\vert a_{0}\right\rangle -c_{\downarrow }\left\vert \downarrow
\right\rangle \left\vert a_{0}\right\rangle  \label{PSI0a}
\end{equation}%
Evolution of the system $\Psi _{0}^{\prime }\longrightarrow \Psi _{2}$ leads
to the same outcome as considered previously $\Psi _{0}\longrightarrow \Psi
_{2}$: different initial conditions $\Psi _{0}$ in (\ref{PSI0}) and $\Psi
_{0}^{\prime }$ in (\ref{PSI0a}) result in, effectively, the same subsequent
state $\Psi _{2}$ in (\ref{PSI2}) that corresponds to $\mathbf{\rho }_{2}=%
\mathbf{\rho }_{3}$. From the many-worlds perspective, this indicates a
merger of two different worlds: the world $\mathcal{W}_{0}$ where $\Psi
=\Psi _{0},$ and the world $\mathcal{W}_{0}^{\prime }$ where $\Psi =\Psi
_{0}^{\prime }$ (assuming that the initial state of $\Psi ,$ has not been
recorded anywhere or memorised by anyone despite being created in a lab,
where the states $\Psi _{0}$ and $\Psi _{0}^{\prime }$ correspond to
different positions of the macroscopic experimental controls). In the spirit
of Everett's principles, any increase of entropy, which represents
irreversible loss of information, involves merging of different worlds with
different alternative pasts (in the same way as, splitting the worlds
corresponds to different alternative futures). Hence, a consistent
time-symmetric version of the many-worlds interpretation presumes existence
of both the world mergers, e.g. $\{\mathcal{W}_{0},\mathcal{W}_{0}^{\prime
}\}\longrightarrow \mathcal{W}_{2}$ and the world splits, e.g. $\mathcal{W}%
_{2}\longrightarrow \{\mathcal{W}_{\uparrow },\mathcal{W}_{\downarrow }\}$.
The former is characterised by increases of the entropy $S_{0}%
\longrightarrow S_{2}$, the latter is associated with apparent decreases of
the entropy $S_{2}\longrightarrow S_{4}$ (compensated by at least the same
entropy increase in the apparatus or elsewhere). Since entropy tends to
increase more than decrease locally and cannot decrease globally, we must
expect that world merges dominate world splits. In other words, there are
many possible alternative futures but even more possible alternative pasts
(as discussed in section \ref{SecB} and figure \ref{fig3}, retrodicting is,
typically, more difficult than predicting --- this indicates that more
alternatives exist in the past than in the future). Our perception that the
past is fixed is created by our causality-based intuition; in fact, most of
the past is irreversibly lost. Here, we refer to the whole body of
information about the past, both unimportant and important (note that
retrodiction can be easier than prediction when the sought information is
selective --- see examples 1-3 of section \ref{SecP}). Everett's
interpretation has been long criticised for producing unphysically large
numbers of multiplying and branching worlds, but in fact, variations in
application of Everett's principles can easily produce more merges than
splits.

\include{figstex}

\end{document}

%% file: klim_tit.tex
\title{The direction of time and Boltzmann's time hypothesis}
\author{A.Y. Klimenko \\ The University of Queensland, SoMME, \\
Brisbane Qld 4072 Australia, \\ email: klimenko@mech.uq.edu.au}

\date{Preprint: March, 2018 \\ (published:  Phys. Scr. 94, 2019, 034002) }

\maketitle

\begin{abstract}
This work explores Boltzmann's time hypothesis, which associates the perceived direction of "time flow" with the second law of thermodynamics. We discuss mechanisms that can be responsible for the action of the second law, for directional properties of time and, ultimately, for the perception that past events cause future events.  Special attention is paid to possibility of testing these mechanisms in experiments. It is argued that CP-violations known in particle physics may offer such an opportunity. 
A time-symmetric version of Everett's many-worlds interpretation (which is based on a thermodynamic interpretation of quantum collapses) is given in the Appendix.  
\end{abstract}

%% file: figstex.tex
\begin{figure}[h]
\begin{center}
\includegraphics[width=12cm,page=1,trim=5cm 9cm 5cm 4cm, clip ]{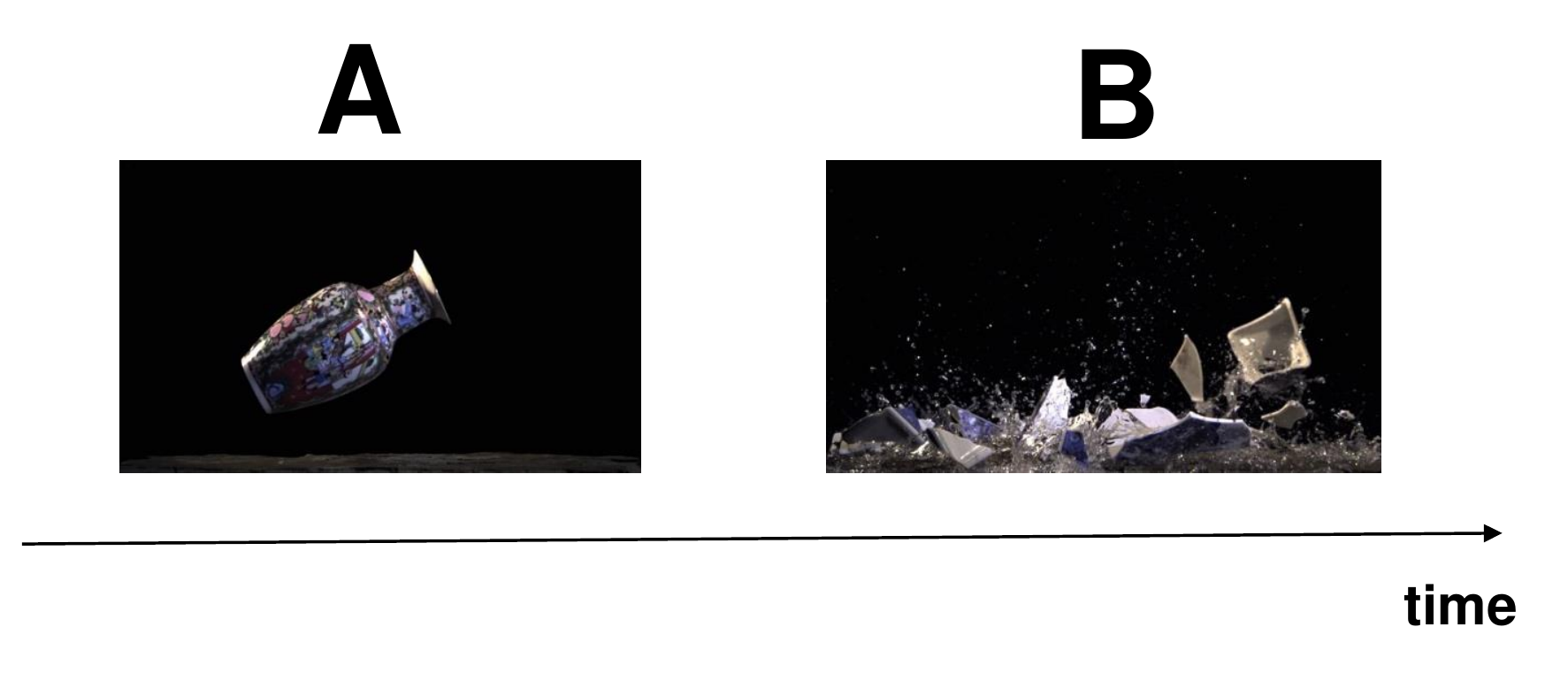}
\caption{Casual and temporal precedence of two events.}
\label{fig1}
\end{center}
\end{figure}

\begin{figure}[h]
\begin{center}
\includegraphics[width=12cm,page=2,trim=1cm 3cm 1cm 4cm, clip ]{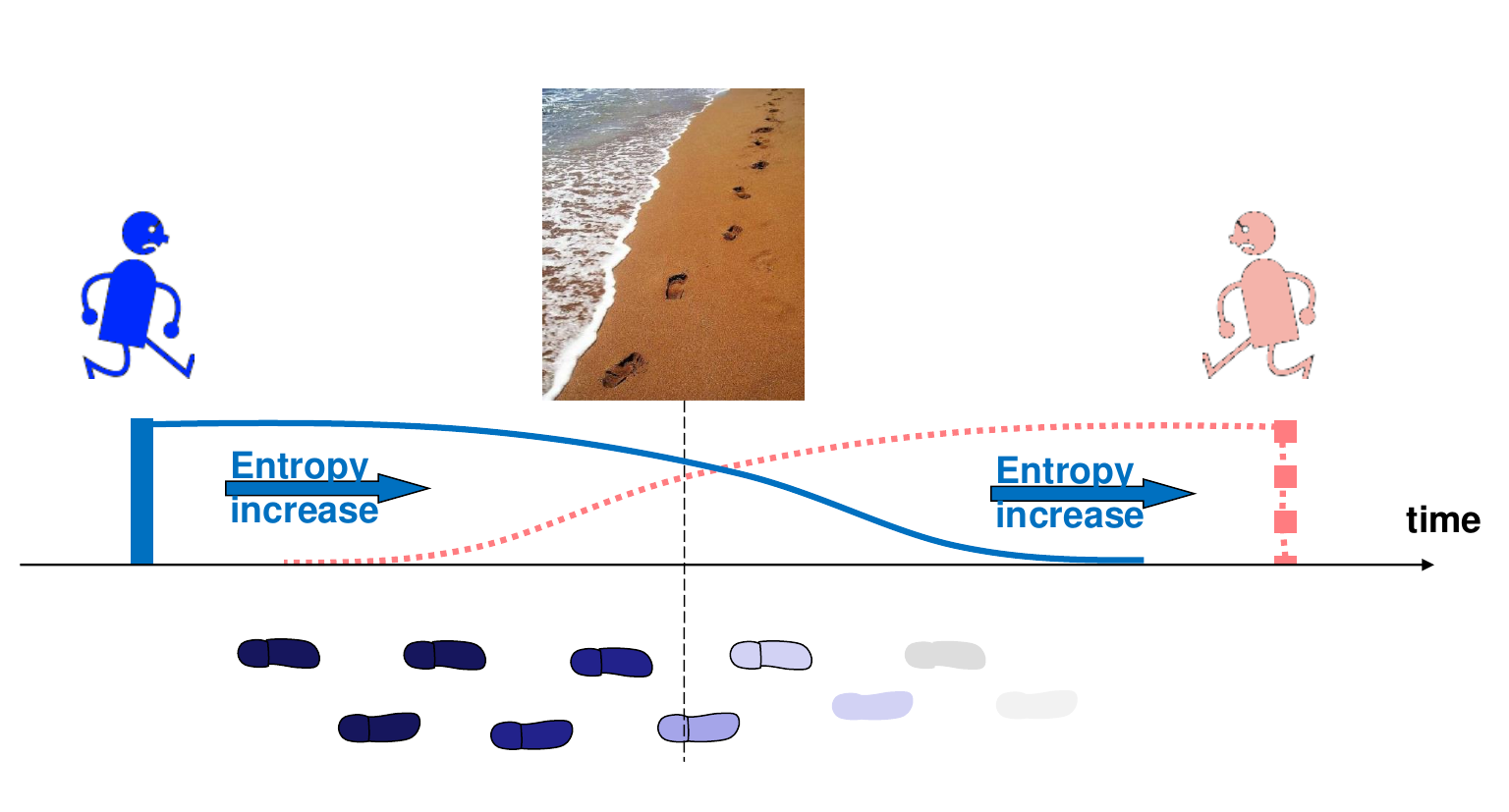}
\caption{Causality, entropy increase and footsteps on a beach. The lines indicate the magnitude of the footprints: solid lines correspond to a realistic case, dashed lines violate the second law of thermodynamics and are not realistic.}
\label{fig2}
\end{center}
\end{figure}

\begin{figure}[h]
\begin{center}
\includegraphics[width=10cm,page=3,trim=5cm 2cm 9cm 5cm, clip ]{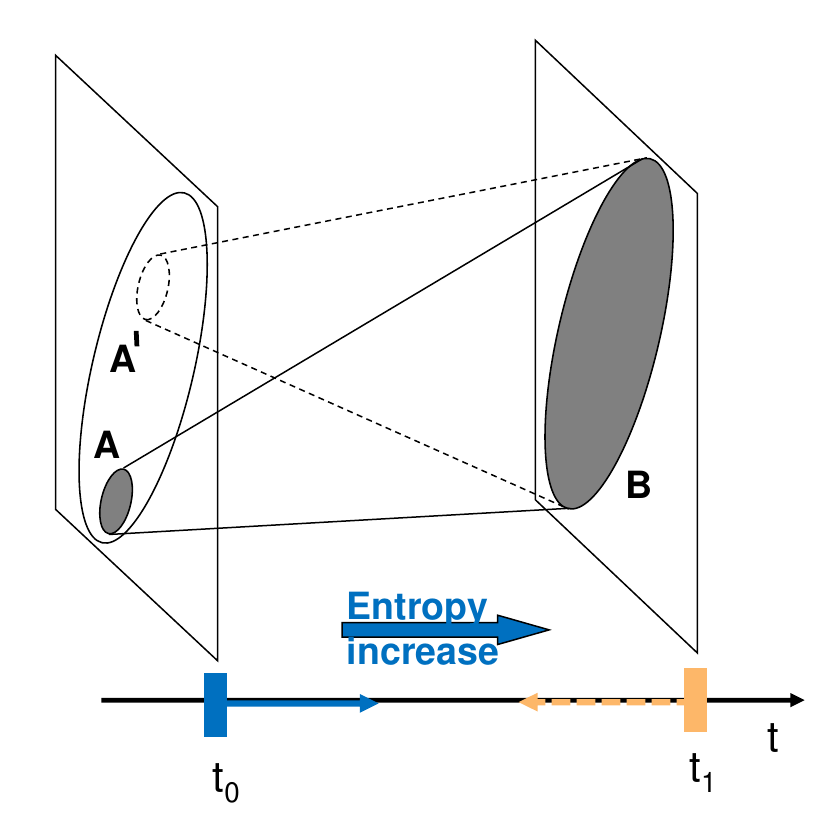}
\caption{Practical causality -- preference for initial (and not final) conditions -- as an effect of the second law of thermodynamics. States A and A$^\prime$ represent alternative initial conditions illustrated as different volumes in the phase space of a dynamic system, while state B, which is close to equilibrium, is a common outcome of different evolutions of the system. The problem is well-posed forward in time and ill-posed backward in time. }
\label{fig3}
\end{center}
\end{figure}

\begin{figure}[h]
\begin{center}
\includegraphics[width=12cm,page=4,trim=4cm 3cm 1.5cm 2cm, clip ]{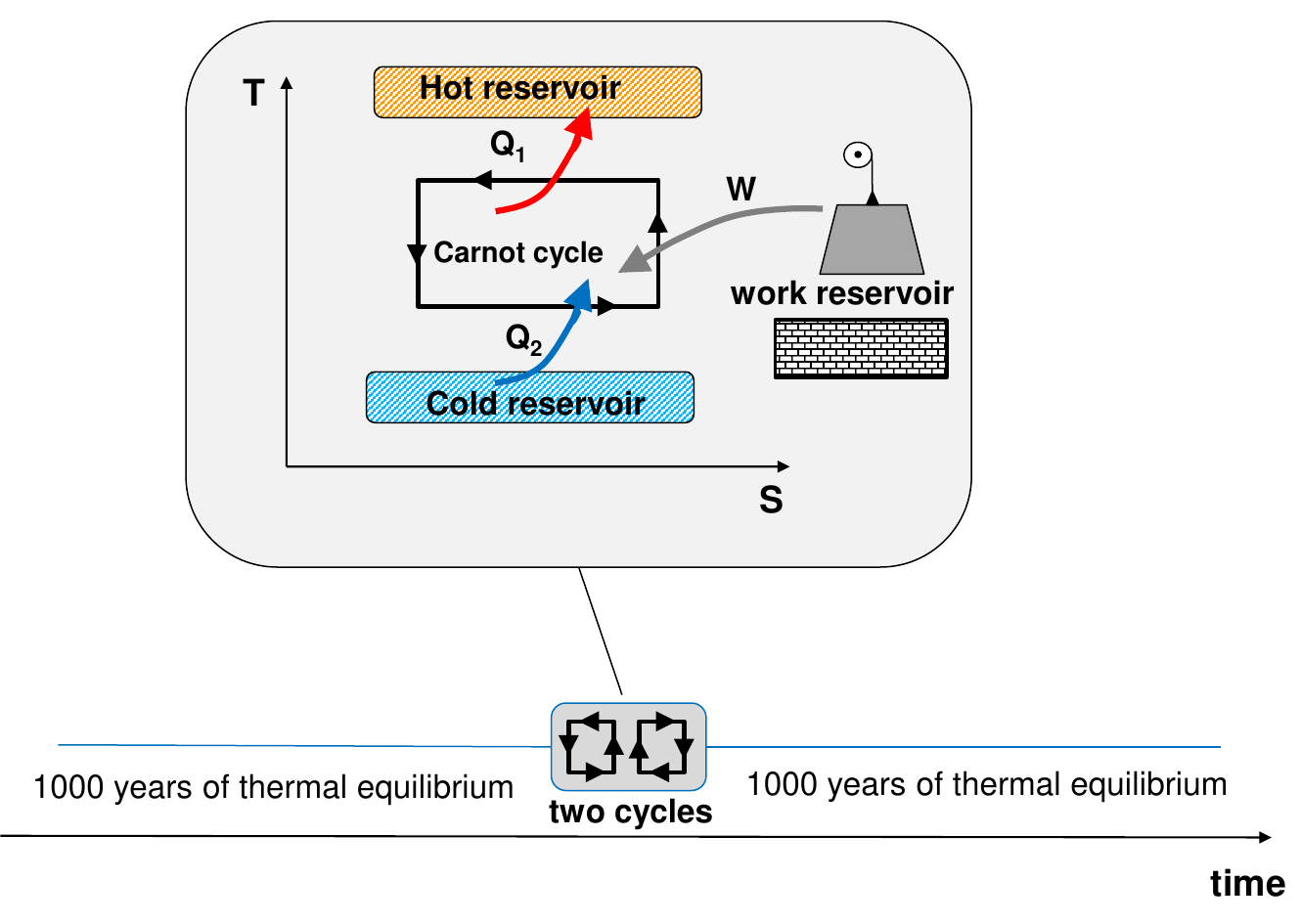}
\caption{Using a Carnot engine to distinguish the directions of time in a remote part of the universe. The two cycles in the middle are screened from any influence of the rest of the universe by a remote location and thousand years of thermal equilibrium before and after activation of the cycles.}
\label{fig4}
\end{center}
\end{figure}

\begin{figure}[h]
\begin{center}
\includegraphics[width=10cm,page=5,trim=2cm 3cm 4cm 2.5cm, clip ]{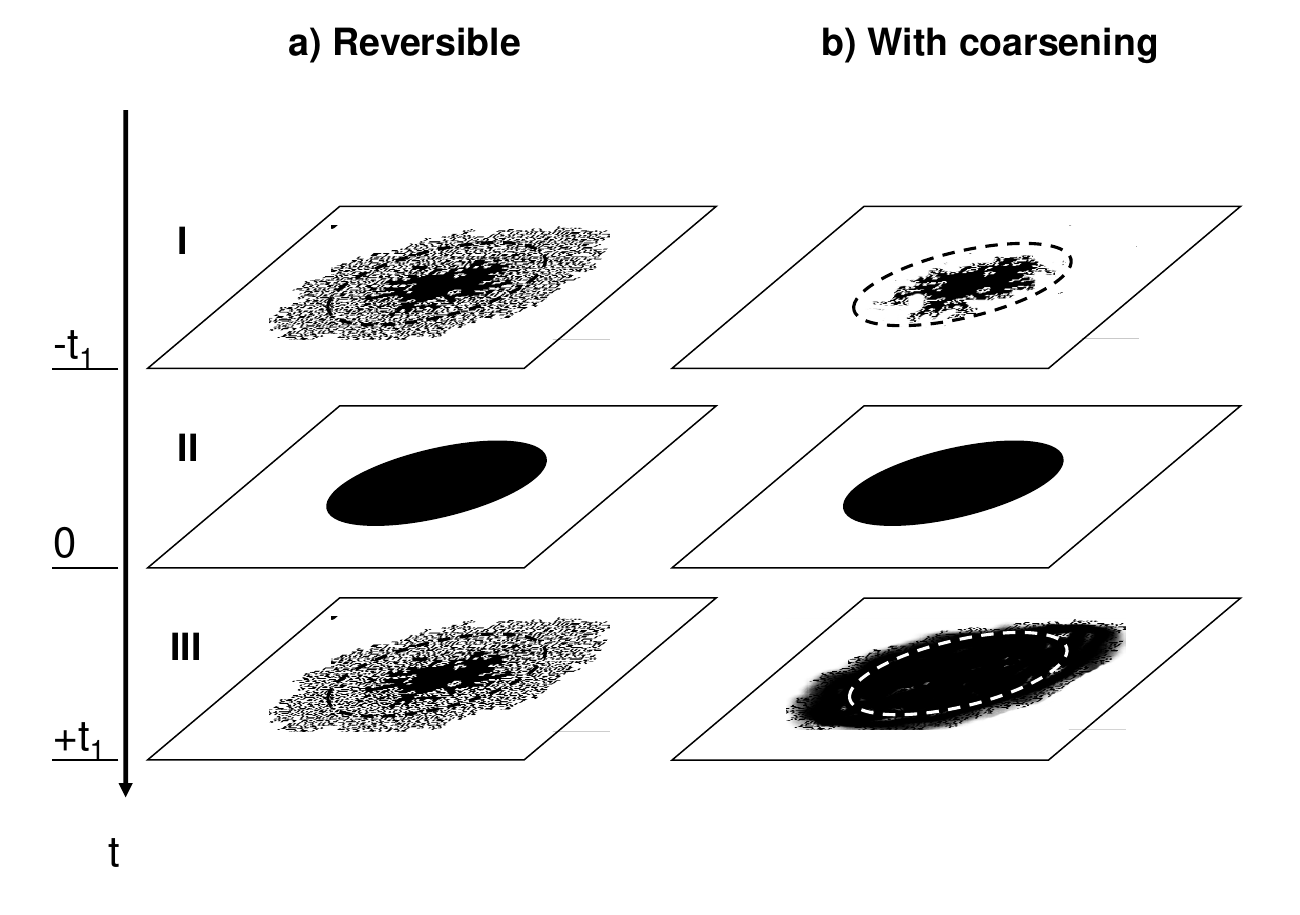}
\caption{Illustration of ergodic mixing in a phase space of a dynamic system of a large dimension: a) without the time primer, the evolution is fully time symmetric and the black domain changes shape but has invariant phase volume associated with constant entropy; b) even tiny, almost undetectable, directional influence of the time primer induces macroscopic changes in the phase volume increasing the entropy forward in time. }
\label{fig5}
\end{center}
\end{figure}

\begin{figure}[h]
\begin{center}
\includegraphics[width=12cm,page=6,trim=2cm 2cm 1cm 2cm, clip ]{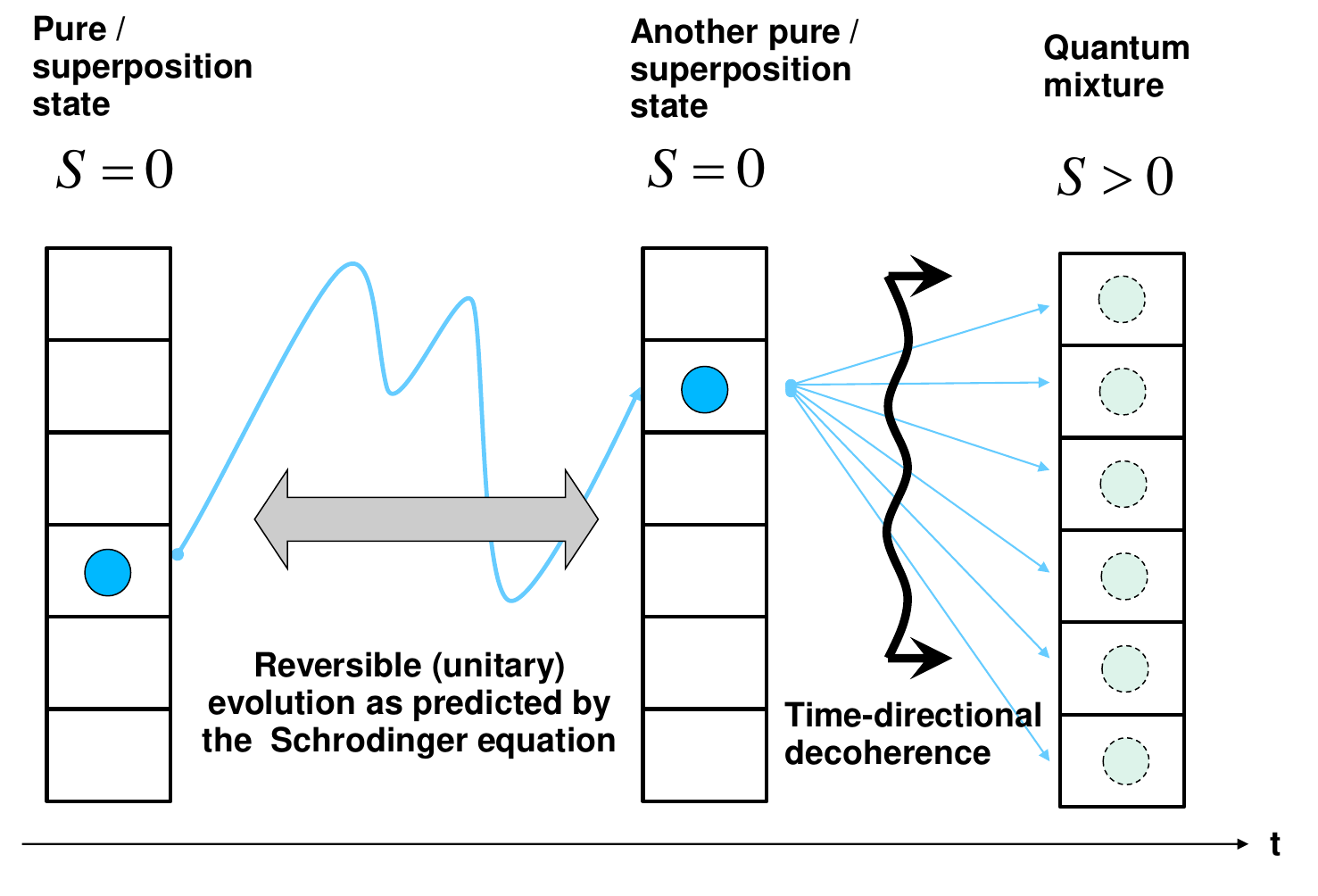}
\caption{Quantum mixing and unitary evolution. Entropy remains constant in unitary evolutions but is increased during decoherence, which converts quantum pure states into quantum mixtures.}
\label{fig6}
\end{center}
\end{figure}

%% file: main.bbl
\begin{thebibliography}{10}

\bibitem{Boltzmann-book}
L.~Boltzmann.
\newblock {\em Lecures on gas thoery}.
\newblock English translation by S.G. Brush. University of California Press,
  Berkeley and L.A., 1964 (1895,1897).

\bibitem{Russell2009}
B.~Russell.
\newblock {\em Our Knowledge of the External World}.
\newblock Taylor and Francis, Florence, 2009.

\bibitem{Reichenbach1971}
H.~Reichenbach.
\newblock {\em The direction of time}.
\newblock University of California Press, Berkeley, 1971.

\bibitem{PriceBook}
H.~Price.
\newblock {\em Time's Arrow and {A}rchimedes' Point: New Directions for the
  Physics of Time}.
\newblock Oxford Univ. Press, Oxford, UK, 1996.

\bibitem{Time1997}
J.~Faye, U.~Scheffler, and M.~Urchs.
\newblock {\em Perspectives on time}.
\newblock Boston studies in the philosophy of science ; v. 189. Kluwer Academic
  Publishers, Dordrecht ; Boston, 1997.

\bibitem{Prig1980}
I.~Prigogine.
\newblock {\em From being to becoming : time and complexity in the physical
  sciences}.
\newblock W. H. Freeman, San Francisco, 1980.

\bibitem{PenroseBook}
R.~Penrose.
\newblock {\em Road to Reality: A Complete Guide to the Laws of the Universe}.
\newblock A. Knopf Inc., 2005.

\bibitem{Zeh2007}
H.~D. Zeh.
\newblock {\em The physical basis of the direction of time}.
\newblock Springer, New York;Berlin;, 5th edition, 2007.

\bibitem{Hawking-time2011}
S.~W. Hawking.
\newblock {\em A brief history of time : from the big bang to black holes}.
\newblock Bantam, London, 2011.

\bibitem{PDG2012}
{Beringer J. et al. (Particle Data Group)}.
\newblock The review of particle physics.
\newblock {\em Phys. Rev.}, D 86:010001, 2012.

\bibitem{Zwart1972}
P.~Zwart.
\newblock The flow of time.
\newblock {\em Synthese}, 24(1):133--158, 1972.

\bibitem{Dowe1992}
P.~Dowe.
\newblock Process causality and asymmetry.
\newblock {\em Erkenntnis}, 37(2):179--196, 1992.

\bibitem{Delayed-choice-2015}
A.~G. Manning, R.~I. Khakimov, R.~G. Dall, and A.~G. Truscott.
\newblock Wheeler's delayed-choice gedanken experiment with a single atom.
\newblock {\em Nature Physics}, 11(7), 2015.

\bibitem{K-PS2012T}
A.~Y. Klimenko.
\newblock What is mixing and can it be complex?
\newblock {\em Physica Scripta}, T155:014047, 2013.

\bibitem{Beretta1991}
E.~P. Gyftopoulos and G.~P. Beretta.
\newblock {\em Thermodynamics. Foundations and Applications}.
\newblock Macmillan, N.Y., USA, 1991.

\bibitem{Abe2009}
S.~Abe.
\newblock Generalized molecular chaos hypothesis and the h theorem: Problem of
  constraints and amendment of nonextensive statistical mechanics.
\newblock {\em Physical Review E}, 79(4), 2009.

\bibitem{Ergo-theory}
J.~L. Lebowitz and O.~Penrose.
\newblock Modern ergodic theory.
\newblock {\em Physics today}, (Feb.):23--28, 1973.

\bibitem{Pauli1928}
W.~Pauli.
\newblock Uber das h-theorem vom anwachsen der entropie vom standpunkt der
  neuen quantenmechanik.
\newblock In {\em Probleme der Modernen Physik. Arnold Sommerfeld zum 60
  Geburtstage}, pages 30--45. Hirzel, Leipzig, 1928.

\bibitem{SciRep2016}
A.~Y. Klimenko.
\newblock Symmetric and antisymmetric forms of the pauli master equation.
\newblock {\em Scientific Reports (nature.com)}, 6:29942, 2016.

\bibitem{Ent2017}
A.~Y. Klimenko.
\newblock {Kinetics of interactions of matter, antimatter and radiation
  consistent with antisymmetric ({CPT}-invariant) thermodynamics.}
\newblock 2017.

\bibitem{Zurek2003}
W.~H. Zurek.
\newblock Decoherence, einselection, and the quantum origins of the classical.
\newblock {\em Reviews of Modern Physics}, 75(3), 2003.

\bibitem{CT-P2009}
N.~Linden, S.~Popescu, A.~J. Short, and A.~Winter.
\newblock Quantum mechanical evolution towards thermal equilibrium.
\newblock {\em Phys. Rev. E}, 79:061103, 2009.

\bibitem{Yukalov2012}
V.I. Yukalov.
\newblock Equilibration and thermalization in finite quantum systems.
\newblock {\em arXiv:1201.2781}, 2012.

\bibitem{QTreview}
A.~Bassia and G.~Ghirardi.
\newblock Dynamical reduction models.
\newblock {\em Physics Reports.}, 379:257--426, 2003.

\bibitem{Diosi2005}
L.~Diosi.
\newblock Intrinsic time-uncertainties and decoherence: comparison of 4 models.
\newblock {\em Brazilian Journal of Physics}, 35(2a), 2005.

\bibitem{Beretta2015}
S.~Cano-Andrade, G.~P. Beretta, and M.~R. {Von Spakovsky}.
\newblock Steepest-entropy-ascent quantum thermodynamic modeling of decoherence
  in two different microscopic composite systems.
\newblock {\em Physical Review A}, 91(1), January 2015.

\bibitem{Diosi1989}
L.~Diosi.
\newblock Models for universal reduction of macroscopic quantum fluctuations.
\newblock {\em Physical Review A}, 40(3):1165--1174, 1989.

\bibitem{Abe2017}
C.~Ou, R.~V. Chamberlin, and S.~Abe.
\newblock Lindbladian operators, von neumann entropy and energy conservation in
  time-dependent quantum open systems.
\newblock {\em Physica A: Statistical Mechanics and its Applications},
  466:450--454, 2017.

\bibitem{KM-Entropy2014}
A.~Y. Klimenko and U.~Maas.
\newblock One antimatter- two possible thermodynamics.
\newblock {\em Entropy}, 16(3):1191--1210, 2014.

\bibitem{Diosi1987}
L.~Diosi.
\newblock A universal master equation for the gravitational violation of
  quantum mechanics.
\newblock {\em Physics Letters A}, 120(8):377--381, 1987.

\bibitem{Penrose1996}
R.~Penrose.
\newblock On gravity's role in quantum state reduction.
\newblock {\em General Relativity and Gravitation}, 28(5):581--600, 1996.

\bibitem{Penrose2014a}
R.~Penrose.
\newblock On the gravitization of quantum mechanics 1: Quantum state reduction.
\newblock {\em Foundations of Physics}, 44(5):557--575, 2014.

\bibitem{Schlosshauer2007}
M.~A. Schlosshauer.
\newblock {\em Decoherence and the quantum-to-classical transition}.
\newblock The frontiers collection. Springer, Berlin ; London, 2007.

\bibitem{Dec3Exp2003}
I.~Chiorescu, Y.~Nakamura, C.~Harmans, and J.~Mooij.
\newblock Coherent quantum dynamics of a superconducting flux qubit.
\newblock {\em Science}, 299(5614):1869--1871, 2003.

\bibitem{Dec2Exp2004}
L.~Hackermüller, K.~Hornberger, B.~Brezger, A.~Zeilinger, and M.~Arndt.
\newblock Decoherence of matter waves by thermal emission of radiation.
\newblock {\em Nature}, 427(6976), 2004.

\bibitem{Dec1Exp2008}
S.~Deléglise, I.~Dotsenko, C.~Sayrin, J.~Bernu, M.~Brune, J.-M. Raimond, and
  S.~Haroche.
\newblock Reconstruction of non-classical cavity field states with snapshots of
  their decoherence.
\newblock {\em Nature}, 455(7212), 2008.

\bibitem{Dec2Exp2003}
L.~Hackermüller, K.~Hornberger, B.~Brezger, A.~Zeilinger, and M.~Arndt.
\newblock Decoherence in a talbot–lau interferometer: the influence of
  molecular scattering.
\newblock {\em Applied Physics B}, 77(8):781--787, 2003.

\bibitem{Dec4Exp2013}
K.~Saeedi, S.~Simmons, J.~Z. Salvail, P.~Dluhy, H.~Riemann, N.~V. Abrosimov,
  P.~Becker, H.-J. Pohl, J.~J.~L. Morton, and M.~L.~W. Thewalt.
\newblock Room-temperature quantum bit storage exceeding 39 minutes using
  ionized donors in silicon-28.
\newblock {\em Science}, 342(6160):830--833, 2013.

\bibitem{SpontDec2014}
S.~Nimmrichter, K.~Hornberger, and K.~Hammerer.
\newblock Optomechanical sensing of spontaneous wave-function collapse.
\newblock {\em Physical review letters}, 113(2):020405, 2014.

\bibitem{K-PhysA}
A.Y. Klimenko.
\newblock Note on invariant properties of a quantum system placed into
  thermodynamic environment.
\newblock {\em Physica A: Statistical Mechanics and its Applications}, 398:65
  -- 75, 2014.

\bibitem{Barbar2016}
{Lees, J. P. {et. al. (BABAR Collaboration)}}.
\newblock Tests of $cpt$ symmetry in
  ${B}^{0}\text{\ensuremath{-}}{\overline{b}}^{0}$ mixing and in
  ${B}^{0}\ensuremath{\rightarrow}c\overline{c}{K}^{0}$ decays.
\newblock {\em Phys. Rev. D}, 94:011101, 2016.

\bibitem{PhysPhil2009}
K.~Camilleri.
\newblock A history of entanglement: Decoherence and the interpretation
  problem.
\newblock {\em Studies In History and Philosophy of Modern Physics},
  40:290--302, 12 2009.

\bibitem{Stamp2012}
P.~C.~E. Stamp.
\newblock Environmental decoherence versus intrinsic decoherence.
\newblock {\em Philosophical transactions. Series A, Mathematical, physical,
  and engineering sciences}, 370(1975):4429, 2012.

\bibitem{Zurek1982}
W.~H. Zurek.
\newblock Environment-induced superselection rules.
\newblock {\em Phys. Rev. Lett.}, 26(8):1862--1888, 1982.

\bibitem{Joos2003}
E.~Joos, C.~Kiefer, and H.~D. Zeh.
\newblock {\em Decoherence and the Appearance of a Classical World in Quantum
  Theory}.
\newblock Springer Berlin Heidelberg, Berlin, Heidelberg, 2 edition, 2003.

\bibitem{VonNeumann1955}
John Von~Neumann.
\newblock {\em Mathematical foundations of quantum mechanics}.
\newblock Investigations in physics ; 2. Princeton University Press, Princeton
  N.J., 1955.

\bibitem{LL5}
L.~D. Landau and E.~M. Lifshits.
\newblock {\em Course of Theoretical Physics vol.5: Statistical physics}.
\newblock Butterworth-Heinemann, Oxford, 1980.

\bibitem{Everett1957}
H.~Everett.
\newblock {"}relative state{"} formulation of quantum mechanics.
\newblock {\em Reviews of Modern Physics}, 29(3):454--462, 1957.

\bibitem{Dewitt1970}
B.~S. Dewitt.
\newblock Quantum mechanics and reality.
\newblock {\em Physics Today}, 23(9):30--35, 1970.

\end{thebibliography}
